\documentclass[12pt,english]{article}
\usepackage{babel}
\usepackage{inputenc}
\usepackage{amsmath}
\usepackage{amsthm}
\usepackage{amsfonts}
\usepackage[dvips]{epsfig}
\usepackage[dvips]{graphics}
\usepackage[dvips]{graphicx}
\usepackage{indentfirst}
\usepackage{rotating}
\usepackage{geometry}
\usepackage{enumerate}
\usepackage{latexsym}
\usepackage{graphicx,graphics,fancybox,array}
\usepackage[justification=centering]{caption}
\usepackage{subcaption}
\usepackage{multirow}
\newtheorem{theorem}{Theorem}

\newtheorem{lemma}{Lemma}

\DeclareGraphicsExtensions{.jpg,.pdf,.png,.jpg,.eps}

\DeclareGraphicsExtensions{.jpg,.pdf,.png,.bmp,.eps}

\title{ Two maxentropic approaches to determine the probability density of compound risk losses}
\author{Erika Gomes-Gon\c{c}alves$^{a}$, Henryk Gzyl$^{b}$, Silvia Mayoral$^{a}$,\\
$^{a}$ Business Administration, Universidad Carlos III de Madrid,\\
 $^{b}$ Centro de Finanzas, IESA, Caracas}
\begin{document}
\maketitle

\begin{abstract}
Here we present an application of two maxentropic procedures to determine the probability density distribution of  compound sums of random variables, using only a finite number of empirically determined fractional moments.  The two methods are the Standard method of Maximum Entropy (SME), and the method of Maximum Entropy in the Mean (MEM). We shall verify that the reconstructions obtained satisfy a variety of statistical quality criteria, and provide good estimations of VaR and TVaR, which are important measures for risk management purposes. We analyze the performance and robustness of these two procedures in several numerical examples, in which the frequency of losses is Poisson and the individual losses are lognormal random variables. As side product of the work, we obtain a rather accurate description of the density of the compound random variable. This is an extension of a previous application based on the Standard Maximum Entropy approach (SME) where the analytic form of the Laplace transform was available to a case in which only observed or simulated data is used.\\
These approaches are also used to develop a procedure to determine the distribution of the individual losses through the knowledge of the total loss. Then, in the case of having only historical total losses, it is possible to decompound or disaggregate the random sums in its frequency/severity distributions, through a probabilistic inverse problem.\\
\end{abstract}

\section{Introduction}
Both in the insurance and the banking industries it is important to know how to compute the density of a compound random variable describing an accumulated random number of losses. In the banking industry is the first step towards the implementation of the advanced measurement approach to determine regulatory capital, and in the insurance industry it is the first step to determine insurance premia. In both cases, the need to do this type of work is to abide by regulatory requirements.

To be specific, in this work we shall suppose that the frequency of losses in a given period of time are described by a compound random variable of the type $S = \sum_{j\geq 0}^N X_j,$ where $N$ is a Poisson random variable of intensity $\ell$, and   $\{X_j,\;\mbox{for} \; j=1,...,N \}$ denote the individual losses which are independent and identically distributed. This type of problems has been studied for a long time, and a variety of techniques exist for its solution, see for example Panjer (2006), but techniques like the proposed here are not yet widely used.

From an abstract point of view, our implementation of the maxentropic methods fall within the techniques to invert Laplace transform from a few values of the parameter. We have actually tried that in a situation in which the Laplace transform could be determined analytically and its values along the real axis are known.  In Gzyl et. al. (2013) the authors applied the SME approach (along with other methodologies) to find the probability density of a compound random variable, in which the frequency is Poisson and the individual losses were $\Gamma(a,b).$ In this case, the compound density $f_S$ may be approximated to any desired degree and different methods of reconstruction can be compared. Unfortunately, when the individual losses are lognormal, neither the density nor the Laplace transform of the compound variable, are available and we have to use numerical methods. This happens in many cases of practical interest, and our example is typical in this regard.

In particular, when the individual losses follow a lognormal distribution, Laplace transform techniques are hard to implement because, to begin with, the Laplace transform of a lognormal density is unknown. By the way, consider the effort in computing it approximately carried out by Liepnik (1991). This impossibility, and the need to recur to numerical methods from the beginning is why we consider this model to describe individual losses.

Besides that, the lognormal distribution has been frequently used to model amounts of claims in various classes of insurance business and in risk theory to model losses caused by different risk events. The fact that it has a heavy tail is important, because it allow us consider the possibility of describing very large claims, which correspond to losses that threaten the solvency of an insurance company or a bank. This has important implications for the determination of premiums, risk reserves and reinsurance (Crow et. al., 1988).

The starting point for us will be the Laplace transform of $S(N)$
\begin{equation}\label{eq1}
E[e^{-\alpha_i S}] = \psi(\alpha_i) = \int_0^\infty e^{-\alpha_is}dF_S(s),\;\; i = 1,...,K.
\end{equation}
\noindent calculated numerically through a simulated data of compounded lognormal losses, and setting $Y=e^{-S}$ as a variable in the interval $[0,1]$ whose density $f_Y(y)$ will be inferred from fractional moments. To begin with, we think of the previous identity as follows
\begin{equation}\label{eq2}
\psi(\alpha_i)=E[Y^{\alpha}]=\int_0^1y^{\alpha_i}dF_Y(y) ,\;\; i = 1,...,K.
\end{equation}
As the distribution $F_S$ of $S$ has a point mass $e^{-\ell}$ at $S=0,$ in order to relate the  $\psi(\alpha)$ to the density $f_S(s)$ of $S$ or that of $Y,$ we have to condition out the mass at $\{Y=1\}$ or at $\{S=0\}.$ For that we consider the conditional version
\begin{equation}\label{eq3}
E[e^{-\alpha_i S}, \quad S > 0] =  \frac{\psi(\alpha_k)-e^{-\ell}}{1-e^{-\ell}} := \mu(\alpha_i),\;\; i = 1,...,K.
\end{equation}
\noindent  which defines $\mu(\alpha_i)$, that will be the input for the maxentropic methods. Once $f_Y$ has been determined, in order to recover $f_{S(N)}$ we have to apply the change of variables $y=e^{-s}$ to obtain $f^*_{S(N)}(s)=e^{-s}f_Y(e^{-s})$.

The two approaches presented in this paper are also used to develop a procedure to determine the distribution of the individual losses from the knowledge of the total severity. Then, in the case of having only a historical record of the total losses, if a model for the frequency of losses is available (and in our case it is),  it is possible to decompound (or to disaggregate) the distribution of losses and obtain the distribution of individual losses. This could be useful for a risk manager that may want to know the distribution of the individual losses in order to apply any particular corrective loss prevention policy.

The remainder of the paper is organized as follows. We recall briefly the basic details of the SME and MEM methods in Section 2. Additionally, in Section 3 we provide a quick overview of the density evaluation methodology used to verify the quality and robustness of the obtained results. In Section 4, we show the results of the implementation of the SME and MEM approaches to determine the distribution of total losses. At this point, we mention the SME and the MEM methods have been applied successfully in a large variety of problems, see Kapur (1989) or Gzyl and Vel\'asquez (2011) for details and references.

Section 5 is devoted to the computation of two of the most commonly used risk measures, namely the $VaR$ and the $TVaR$ using the maxentropic density as loss probability density. This could be interesting for risk managers, who consider insure the operational risk losses to decrease the capital charges. Section 6 is devoted to the complementary problem of disaggregation which will play the role of check up test for our procedure. In Section 7, we present some concluding remarks and finally, the Appendix in the Section 8 describes in detail a variety of tests and graphical tools used to analyze the quality of the results.

\section{The maxentropic approaches}
Bellow we  review the basis of the SME and MEM methods used to solve the problem of finding the density of the total severity from the knowledge of a small number of fractional moments.

\subsection{The Standard method of Maximum Entropy (SME)}
This is a variational procedure proposed by Jaynes (1957) to solve the (inverse) problem consisting of finding a probability density $f_Y(y)$ (on $[0,1]$ in this case), satisfying the following integral constraints:
\begin{equation}\label{ME}
\int_0^1 y^{\alpha_k} f_Y(y)dy = \mu_Y(\alpha_k)\qquad\mbox{for}\qquad k=0,1,...,K.
\end{equation}
We set $\alpha_0=0$ and $\mu_0=1$ to take care of the natural normalization requirement on $f_Y(y).$ The intuition is rather simple: The class of probability densities satisfying (\ref{ME}) is convex. One can pick up a point in that class one by maximizing (or minimizing) a concave (convex) functional (an ``entropy'') that achieves a maximum (minimum) in that class. That extremal point is the ``maxentropic'' solution to the problem. It actually takes a standard computation to see that when the problem has a solution it is of the type
\begin{equation}\label{sol1}
f_K^{*}(y) = \exp\left(-\sum_{k=0}^K \lambda^{*}_k y^{\alpha_k}\right)
\end{equation}
\noindent in which the number of moments $K$ appears explicitly. It is usually customary to write $e^{-\lambda^{*}_0} = Z(\mbox{\boldmath $\lambda$}^{*})^{-1},$ where $\mbox{\boldmath $\lambda^{*}$} = (\lambda_1^{*},...,\lambda_K^{*})$ is a $K-$dimensional vector.
Clearly, the generic form of the normalization factor is given by
\begin{equation}\label{zeta}
Z(\mbox{\boldmath $\lambda$}) = \int_0^1 e^{-\sum_{k=1}^K \lambda_k y^{\alpha_k}}dy.
\end{equation}
 With this notation the generic form of the solution looks like
\begin{equation}\label{sol2}
f_K^{*}(y) = \frac{1}{Z(\mbox{\boldmath $\lambda$}^{*})}e^{-\sum_{k=1}^K \lambda^{*}_k y^{\alpha_k}} = e^{-\sum_{k=0}^K \lambda^{*}_k y^{\alpha_k}}.
\end{equation}
To complete, it remains to specify how the vector $\mbox{\boldmath $\lambda$}^{*}$ can be found. For that one has to minimize the dual entropy:
\begin{equation}\label{dual}
  \Sigma(\mbox{\boldmath $\lambda$},\mbox{\boldmath $\mu$}) = \ln Z(\mbox{\boldmath $\lambda$}) + \langle\mbox{\boldmath $\lambda$},\mbox{\boldmath $\mu$}_Y\rangle
\end{equation}
\noindent where $\langle\mathbf{a},\mathbf{b}\rangle$ denotes the standard Euclidean scalar product and $\mbox{\boldmath $\mu$}$ is the $K-$vector with components $\mu_k,$ and obviously, the dependence on $\mbox{\boldmath $\alpha$}$ is through $\mbox{\boldmath $\mu$}_Y.$

\subsection{The method of Maximum Entropy in the Mean (MEM)}
The MEM provides another interesting approach to solve the problem of determining $f_Y(y)$ such that Equation (\ref{eq2}) holds. It can be summed up by saying that it consists in a technique to obtain $f_Y(y)$ or its discretized version, as the expected value of an auxiliary probability distribution which is determined by an entropy maximization procedure. To implement MEM numerically, the first step consists of discretizing the problem. This leads to a system of equations like:
\begin{equation}\label{Eq1}
\sum_{j=1}^M A_{i,j}x_j = \mu_i, i=0,1,...,K\;\;\mbox{with}\;\;x_j \geq 0\;\;\mbox{and}\;\;j=1,...,M.
\end{equation}

\noindent Here we have set $x_j=(1/N)f((j-1)/M)$ and $ A_{i,j} = (\frac{2j-1}{2M})^{\alpha_i},$ for j=1,...,M. The first factor in front of the definition of $x_j$ comes from the discretization $dy\approx 1/M,$  and $A_{i,j}$ is obtained as the midpoint approximation of $y^{\alpha_i}$ for the chosen partition. We have added a normalization constraint by choosing $\alpha_0=0.$ Note that with this choice we have $A_{0,j}=1$ for $j=1,...,M$ and $\sum_{j=0}^Mx_j= \mu_0=1$ as normalization condition for $\alpha_0=0.$ Observe that if we consider a partition of size $200$, as we have $K+1$ moments, we have a system of $K+1$ equations to determine $200$ unknowns subject to a positivity constraint. Actually, the original problem consists of $K+1$ equations to determine a continuous function, so the discretization may seem to be an improvement from the dimensionality point of view.

To continue, to take care of the positivity constraint, we consider a space $\Omega=[0,\infty)^M$ with its Borel sets $\mathcal{F}.$  Denote by $\mbox{\boldmath $\xi$} =(\xi_1,...\xi_M)$ a generic point in
$\Omega$ and define the coordinate maps $X_j(\mbox{\boldmath $\xi$}):\Omega \rightarrow [0,\infty)$  by
$X_j(\mbox{\boldmath $\xi$})=\xi_j.$ On $(\Omega,\mathcal{F})$ we place a reference measure
$dQ(\mbox{\boldmath $\xi$})$ and we search for a measure $P<<Q$ such that
\begin{equation}\label{mem1}
\sum_{j=1}^{M}A_{i,j}E_P[X_j] = \mu_i\;\;\mbox{for}\;\;0=1,...,K.
\end{equation}
The choice of the measure $Q$ is up to the modeler, and it may be thought of as the first guess of $P.$ The only restriction upon it is that the convex hull generated by its support is $\Omega.$ The purpose of that condition is to ensure that any strictly positive density $\rho(\mbox{\boldmath $\xi$})$ of $P$ respect to $Q$ is such that
$$E_P[X_j] = \int_{\Omega}\xi_j\rho(\mbox{\boldmath $\xi$})d\mbox{\boldmath $\xi$} \in [0,\infty),$$
\noindent that is, the positivity constraint is automatically satisfied. The other constraints will be achieved by a special choice of $P.$ With the notations introduced above, we note that the class of probabilities
$$\mathcal{P} = \{P << Q\;\;\mbox{such that (\ref{mem1}) holds true}\}$$
is a convex, closed set if not empty, which we suppose to be the case. On this set we define the entropy function
$$S_Q(P) = -\int_{\Omega} \rho(\mbox{\boldmath $\xi$})\ln\left(\rho(\mbox{\boldmath $\xi$})\right)dQ(\mbox{\boldmath $\xi$})$$
\noindent whenever the integral is finite or $\infty$ otherwise.
\begin{equation}\label{probmem}
\mbox{Find}\;\;P^* \in \mathcal{P}\;\;\mbox{which maximizes}\;\;S_Q(P).
\end{equation}
From now on the routine is pretty much as in the previous Section. It is clear that the MEM uses the SME as a stepping stone. The problem is similar but in another setup. The generic solution is of the type

\begin{equation}\label{3}
\rho^*(\mbox{\boldmath $\xi$}) = \frac{1}{Z(\mbox{\boldmath $\lambda$}^*)}e^{-\left(\langle\left(\mbox{\boldmath $A^t\lambda$}^*\right),\mbox{\boldmath $\xi$}\rangle\right)}
\end{equation}
\noindent where, again $\mbox{\boldmath $\lambda$}^*$ is obtained minimizing the dual entropy function

\begin{equation}\label{dual2}
\Sigma(\mbox{\boldmath $\lambda$},\mbox{\boldmath $\mu$}) = Z(\mbox{\boldmath $\lambda$}) + \langle\mbox{\boldmath $\lambda$},\mbox{\boldmath $\mu$}\rangle,
\end{equation}
where, recall, $\mbox{\boldmath $\lambda$}$ is a $K$dimensional vector, and $\mbox{\boldmath $\mu$}$ is a $K$dimensional vector of constraints defined (\ref{Eq1}). This time, the function $Z(\mbox{\boldmath $\lambda$})$ is defined by
$$Z(\mbox{\boldmath $\lambda$}) = \int_{\Omega} e^{-\left(\langle\left(\mbox{\boldmath $A^t\lambda$}^*\right),\mbox{\boldmath $\xi$}\rangle\right)}dQ(\mbox{\boldmath $\xi$}).$$
The following result is a simplified version of the duality theory presented in Chapter 4 of Borwein and Lewis (2000). It provides a sound basis for the computations corresponding to the two example presented below.

\begin{theorem}
Let us suppose that $\inf \Sigma(\mbox{\boldmath $\lambda$},\mbox{\boldmath $\mu$})$ is achieved at an interior point $\mbox{\boldmath $\lambda$}^*$ of $\{\mbox{\boldmath $\lambda$} \in \mathbb{R}^K | Z(\mbox{\boldmath $\lambda$}) < \infty \}.$ In this case, the probability $P^*$ solving (\ref{probmem}) has density $\rho^*(\mbox{\boldmath $\xi$})$ given by (\ref{3}), and
$$S_Q(P^*) = \Sigma(\mbox{\boldmath $\lambda$}^*,\mbox{\boldmath $\mu$}).$$
\end{theorem}

 In the next section we shall illustrate the MEM in one specific setup. The basic idea is that the $x_j$ are to be estimated as expected values with respect to a probability $P$ to be determined as explained above.

\subsubsection{Poisson reference measure}
As reference measure we use a product of Poisson measures, i.e., we take
$$q(d\xi) = e^{-\eta}\sum_{k\geq 0} \frac{\eta^k}{k!}\epsilon_{\{k\}}(d\xi).$$
Here we use $\epsilon_{\{a\}}(d\xi)$ to denote the unit point mass (Dirac delta) at $a.$ Certainly the convex hull of the non-negative integers is $[0,\infty).$ Notice that now
$$Z(\mbox{\boldmath $\lambda$}) = \prod_{j\geq 0}^M\exp\Big(-\eta\Big(1- e^{-(\mbox{\boldmath $A^t\lambda$})_j}\Big)\Big)$$
from which we obtain
$$\Sigma(\mbox{\boldmath $\lambda$}) = -\eta\sum_{j-1}^M\Big(1- e^{-(\mbox{\boldmath $A^t\lambda$})_j}\Big) + \mbox{\boldmath $<\lambda,m>$} $$
Notice now that if $\mbox{\boldmath $\lambda$}^*$ minimizes that expression, then the estimated solution to (\ref{Eq1}) is
\begin{equation}\label{repsol2}
x^*_j = e^{-(\mbox{\boldmath $A^t\lambda$}^*)_j}
\end{equation}

Do not forget that above, $\mbox{\boldmath $A^t\lambda$}_j = \sum_{i=0}^8\lambda_iA_{i,j}.$ Recall as well that $A_{0,j}=1$ for $j=1,...,M.$ As $\sum_{j=0}^M x^*(j)=1,$ we can rewrite (\ref{repsol2}) as
\begin{equation}
\label{repsol3}
x^*_j = \frac{e^{-(\mbox{\boldmath $\hat{A}^t\hat{\lambda}$}^*)_j}}{z(\mbox{\boldmath $\hat{\lambda}$}^*)}
\end{equation}
\noindent where we redefined $\hat{\mathbf{A}}$ as the matrix obtained from $\mathbf{A}$ by deleting the zero-th row, and $\mbox{\boldmath $\hat{\lambda}$}$ as the {$8-$dimensional  vector obtained by deleting $\lambda_0.$ To complete,
$$z(\mbox{\boldmath $\hat{\lambda}$}^*)=e^{-\lambda_0}= \sum_{j=1}^M e^{-\big(\mbox{\boldmath $\hat{A}^t\hat{\lambda}$}^*\big)_j}.$$
Recall that $x^*_j=(1/M)f_Y^*((j-1)/M),$ from which we determine $f^*_Y(y)$ approximately by interpolation.
%The results are very sensitive to the interpolation scheme

\section{Quality of the reconstructions}
Once a density has been determined, it is necessary to  test whether it is consistent with the data. The evaluation process is inherently a statistical problem, which involves exploring, describing, and making inferences about data sets containing observed and estimated values. Here we describe a variety of tests, and in the Appendix we add further detail about them.

Exploratory tests to asses the quality of a reconstruction, include visual comparisons through the use of graphical tools like reliability and calibration plots which measures the agreement between the estimation and the observed data. The reliability plot, popularly known as QQ-plot serves to determine the quality of a fit by the proximity of the quantiles of the maxentropic distribution to the diagonal, the closer the better the approximation. A similar tool is the marginal calibration plot which is the graph of  $F^*_S(s_j) - F_n(s_j)$ versus the total losses $\{s_j| j=1, ..., n\}$, where $F^*_S(\cdot)$ is the cumulative distribution function of the reconstructed maxentropic density, and $F_n(\cdot)$ is the observed or empirical distribution function of the losses $S.$ Here, minor fluctuations about zero means that the observations and the maxentropic estimations have the same (or nearly the same) marginal distribution (See Appendix).

Numerical comparisons are also considered for the evaluation of the results. Among them, we compute the  $L_1$ and $L_2$  distances between the densities and the
histogram of the observed data. These distances are calculated by

$$L_1=\sum_{k=0}^{G-1} \int_{b_k}^{b_{k+1}} |f^*_S(s)-f_n(s)|ds + \int_{b_G}^{\infty}|f^*_S(s)|ds $$

$$L_2=\sqrt{\sum_{k=0}^{G-1} \int_{b_k}^{b_{k+1}} (f^*_S(s)-f_n(s))^2ds + \int_{b_G}^{\infty}(f^*_S(s))^2ds }$$

\noindent
where $b_k$ and $b_{k+1}$ are the boundaries of the bins in the histogram, $G$ is the number of partitions or bins, $f^*_S$ is the maxentropic (reconstructed) density and $f_n$ is the density obtained from the histogram (i.e. (frequency in the bin $k$)/(size of the data set)). This measure has the disadvantage of depending on the location and the number of bins of the histogram.

Also, we consider the Mean Absolute Error (MAE) and the Root Mean Squared Error (RMSE) as measures of error between the distribution functions obtained with the maxentropic methods versus the observed data. These are calculated as follows:

 $$MAE=\frac{1}{n} \sum\limits_{j=1}^n |F^*_S(s_j)-F_n(s_j)|$$

 $$RMSE=\sqrt{\frac{1}{n} \sum\limits_{j=1}^n  (F^*_S(s_j)-F_n(s_j))^2}$$

\noindent where $F^*_S(s_j)$ is the distribution function of the maxentropic procedure, and
$F_n(s_j)$ is the distribution function of the observed data (Hyndman et. al., 2006). This measures of fit have the advantage of not relying on the number of bins, due to $F^*_S(\cdot)$ and $F_n(\cdot)$ can be calculate for all the points of the sample.

Another possibility, is to test whether the estimated density function $f^*_S(\cdot)$, equals the true underlying density function $f_S(\cdot).$ Another way to do this, is based on an integral transform that dates back to Rosenblatt (1952), and popularized by Diebold et al. (1998), which consists in testing whether the probability integral transforms (PIT) is independent and uniformly distributed.

The probability integral transform (PIT) of the data is defined by
\begin{equation}\label{pt}
\int^{s_j}_{-\infty} f^*_S(s)ds=F^*_S(s_j)=p_j
\end{equation}
\noindent where $s_j$ is the $j-$th element of the sample of interest, in this case, the observed (simulated) total loss. That is,  $s_j$ are sampled from a population described by an unknown density $f_S(\cdot).$ In the definition $f^*_S(\cdot)$  is the reconstructed (maxentropic) density, and $p_j$ is the probability integral transform (PIT) of $s_j$, which should follow an i.i.d uniform distribution ($p_j \sim Unif(0,1)$) if $f^*_S$ happened to equal $f_S.$

Deviations from uniformity will indicate that the reconstruction may have failed to capture some aspect of the underlying data generation process. To test uniformity and independence in the PIT test, a visual inspection of a PIT-histogram and autocorrelation plots is used along with additional tests like the KS-test, the Anderson-Darling test, and the Cram\'{e}r-Von Mises test (Tay et. al, 2000). Additionally, we also consider the Berkowitz back test approach, which consists of taking the inverse normal transformation of the PIT and then applying a joint test for normality and independence, that is sometimes combined with the normality test of Jarque-Bera.

The calculation of the VaR and the TVaR measures using the SME and MEM densities, may be considered to be an additional way to evaluate the quality of the reconstruction. This is done by comparing these values with the empirical VaR and TVaR obtained from the observed data.

Many of the comparisons that we perform in this paper are made with respect to a simulated compound random variable whose probability density is unknown, but the sample is large enough to provide us with a good approximation of the data used as input to the maxentropic methods. Throughout the paper we will refer to this sample as ``the observed data''.

On the other hand, we also perform comparisons using a sample that we call ``the test set'', which is an independent data set that comes from the same population as the observed data, and which is used to audit the results. This is done in order to avoid the problem of overfitting,  which is defined as the situation in which the model gives better results for the ``observed'' set than for other data set which comes from the same population. Besides this helps us to evaluate the ability of the maxentropic density to perform well on unobserved data.

\section{Numerical essays}

The examples that we consider consists of a compound process in which the frequency of events during a given period of time is described by a Poisson random variable $N$ of parameter $\ell,$  and the individual losses $X_j$ for $j\geq 1,$ are distributed according a lognormal distribution ($X \sim logN(\mu,\sigma^2)$). In this case the Laplace transform of $S(N)$ cannot be computed analytically, but it can be estimated numerically.

We proceed as follows:
\begin{enumerate}
  \item We generate a sample of size 8.000 from a compound distribution $S(N)=\sum\limits_{1 \leq j}^{N} X_j$, where $\{N \geq 0 | \quad N=n_1, n_2, \ldots, n_n\}$,  are the sizes of the lognormal samples
  with mean $\mu$ and standard deviation $\sigma$, and $n$ is the total number of poisson samples $n_k$ of parameter $\ell$.
 Using the simulated losses, we calculate the moments $\mu(\alpha_i),$ which are the input data needed to obtain the maxentropic distributions.
  \item We apply the SME and MEM approaches described in Sections (2.1) and (2.2), using 8 fractional moments $\mu(\alpha_i)$ of the exponential of the variable, where $\alpha_i=\frac{1.5}{i}$ with $i=1,\ldots, 8$. That is $K=8$ is the number of non zero moments. The minimization procedure is carried out using the Barzilai-Borwein's (1988) algorithm.
  \item Each approach is implemented in five cases. One is described with full detail and called Case (1),  characterized by the parameters $\ell=3$, $\mu=0$, $\sigma=0.25,$ and the other four (Cases 2-5) are used to verify the robustness of the maxentropic approaches with respect to changes in the parameters. Thus, we consider two cases with the same lognormal parameters  $\mu=0,\sigma=0.25$, but different Poisson parameters (Cases 2-3) and other two cases with the same Poisson parameter $\ell=3$ and different lognormal parameters (Cases 4-5).
  \item Once the maxentropic densities have been obtained, we evaluate the quality of the reconstructions using a variety of criteria (described in  Section (3) and in the Appendix) for the two independent data sets, namely, the observed data set of size 8000 and a test data set of size 1500.
\end{enumerate}

\subsection{Numerical Reconstructions with the SME approach}

We start with a compound process where the frequency $N$ is a Poisson variable of parameter $\ell=3$ and the individual losses are described by a lognormal distribution with parameters $\mu=0$, $\sigma=0.25$, the resulting compound sum is called $S,$ which represents the total losses that can affect a business.

In Figure (\ref{SME_results}) we display the density, the distribution function, the marginal calibration and reliability diagram of the SME based reconstruction. These plots allow us to observe the performance of the SME method with respect to the empirical histogram and the empirical (cumulative) distribution function, by using different perspectives of the same result which gives us a visual idea of how good the obtained reconstruction is.

\begin{figure}[h!]
        \centering
        \begin{subfigure}[b]{0.4\textwidth}
                \includegraphics[width=\textwidth]{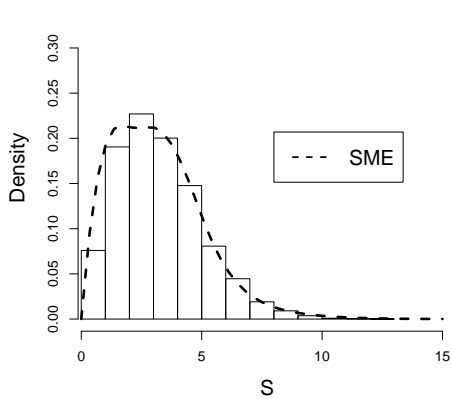}
                \caption{SME density}
                \label{fig:ex4}
        \end{subfigure}
        \begin{subfigure}[b]{0.4\textwidth}
                \includegraphics[width=\textwidth]{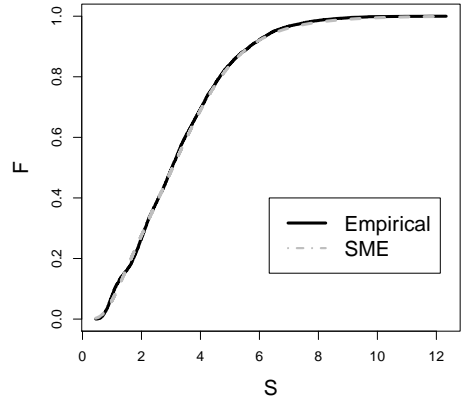}
                \caption{SME distribution function}
                \label{fig:Fex4}
        \end{subfigure}
        \begin{subfigure}[b]{0.4\textwidth}
                \includegraphics[width=\textwidth]{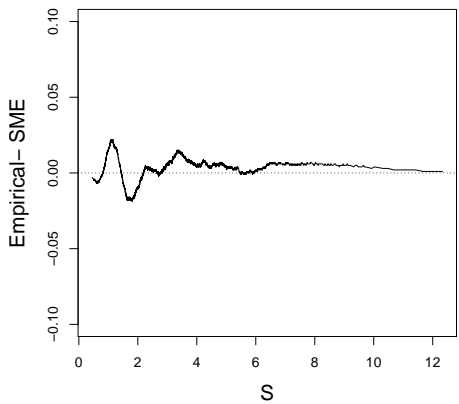}
                \caption{SME Marginal calibration}
                \label{fig:marginal_ex4}
        \end{subfigure}
        \begin{subfigure}[b]{0.4\textwidth}
                %\addtocounter{subfigure}{4}
                \includegraphics[width=\textwidth]{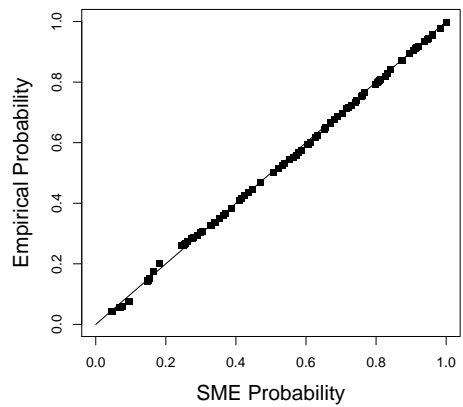}
                \caption{SME Reliability Diagram}
                \label{fig:reliability_ex4}
        \end{subfigure}
        \caption{SME results of the Case (1): $\ell=3$, $\mu=0$, $\sigma=0.25$}
        \label{SME_results}
\end{figure}

Notice in Figure (\ref{fig:ex4}) that the SME density and the histogram of the observed data appear to be close. The same can be seen in Figure (\ref{fig:Fex4}) for the  (cumulative) distribution function of the SME and the observed data, whose differences seem to be imperceptible. In Figure (\ref{fig:marginal_ex4}) we display the marginal calibration diagram that allows us to observe the differences between the (cumulative) distribution functions of the SME reconstruction and the observed data. This figure shows that the differences are not larger than 0.022. Such small fluctuations about zero are indicators of the good quality of the reconstruction. Additionally, in Figure (\ref{fig:reliability_ex4}) we have the reliability diagram of the observed frequency against the SME density. The diagram measures the agreement between estimated probabilities and the observed frequencies, indicated by the proximity of the plotted curve to the diagonal. Here we can see that there is only a very small deviation at the beginning of the graph.

Even though the results of the Figure (\ref{SME_results}) seems to indicate a good reconstruction, as measure of closeness we also consider the $L_1$ and $L_2$ norms of the distances between the reconstructed density and the empirical density, as well as the MAE and RMSE errors detailed in Section (3). In Table (\ref{tab:error_case_a}) we display the results of the computations. These values confirm the plausibility of the good quality of the reconstruction displayed in Figure (\ref{SME_results}). The MAE and RMSE distances with respect to the histogram are reasonably small, of order $10^{-3},$ and the  $L_1$ and $L_2$ distances between the maxentropic and the histogram are good enough
\begin{table}[h!]
\small
\centering
\begin{tabular}{|c|c|c|c|c|c|c|}
 \multicolumn{5}{c}{}\\
\hline
    Approach &  $L_1$     &   $L_2$        & MAE      & RMSE                    \\ \hline
     SME     & 0.1225  & 0.0598      & 0.0071   & 0.0089       \\
  \hline
\end{tabular}
\caption{Errors SME approach, \\ Case (1): $\ell=3$, $\mu=0$, $\sigma=0.25$}
\label{tab:error_case_a}
\end{table}

It is convenient to test density reconstructions for correct calibration. This consists of testing whether the inverse probability transforms (PIT) is independent and uniformly distributed. Deviations from uniformity may indicate a poor reconstruction. In Figure (\ref{PIT(ex4)}) we display the PIT transformation and correlograms of different powers for Case (1). As we can see, the PIT histogram seems to be uniform and the correlation plots do not show any sign of dependence.

\begin{figure}[h!]
  \centering
  \includegraphics[width=9cm,height=7cm]{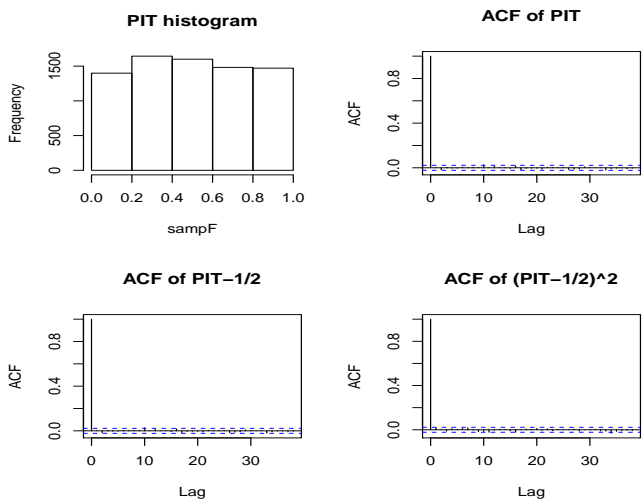}
  \caption{Probability integral transform (PIT) histogram and sample autocorrelation functions
for the SME approach. Case (1) $\ell=3$, $\mu=0$, $\sigma=0.25$.}
  \label{PIT(ex4)}
\end{figure}

As said, to test robustness of the SME method, we performed reconstructions with simulated data for other values of the parameters. In each of the cases we went through the same routine: we generated the data, computed the moments and carried out the maxentropic procedure. In Figure (\ref{SME_all_cases}) we display the different SME reconstructions, along with the histogram of the observed data.  A glance at the figures should convince us that the different reconstructions seem to fit the observed data reasonably well. The same criteria that we used above to measure quality of the Case (1) yield consistent results in these cases as well.

\begin{figure}[h!]
        \centering
        \begin{subfigure}[b]{0.4\textwidth}
                \includegraphics[width=\textwidth]{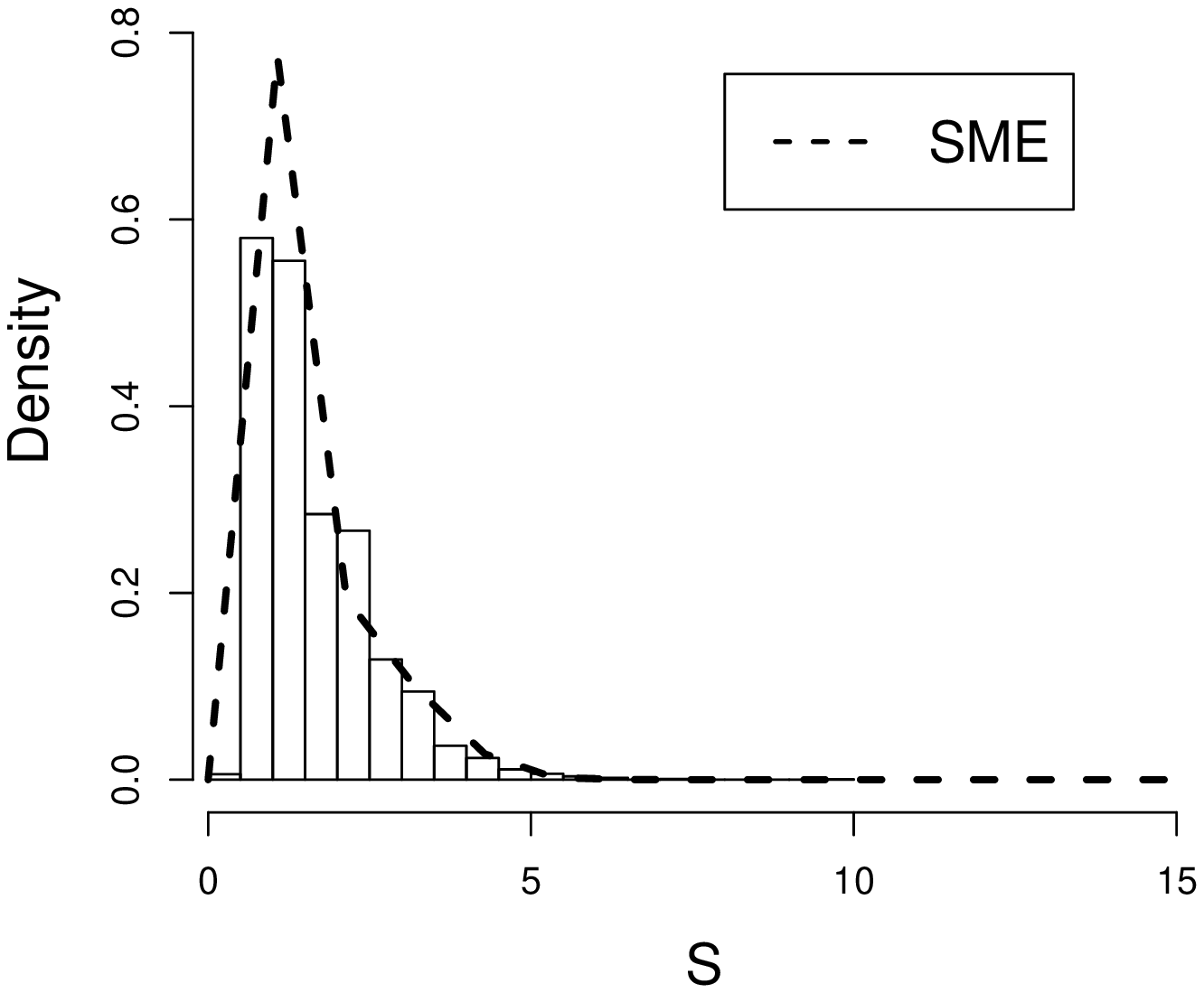}
                \caption{Case (2): \\ $\ell=1$, $\mu=0$, $\sigma=0.25$}
                \label{fig:ex2}
        \end{subfigure}%
        \begin{subfigure}[b]{0.4\textwidth}
                \includegraphics[width=\textwidth]{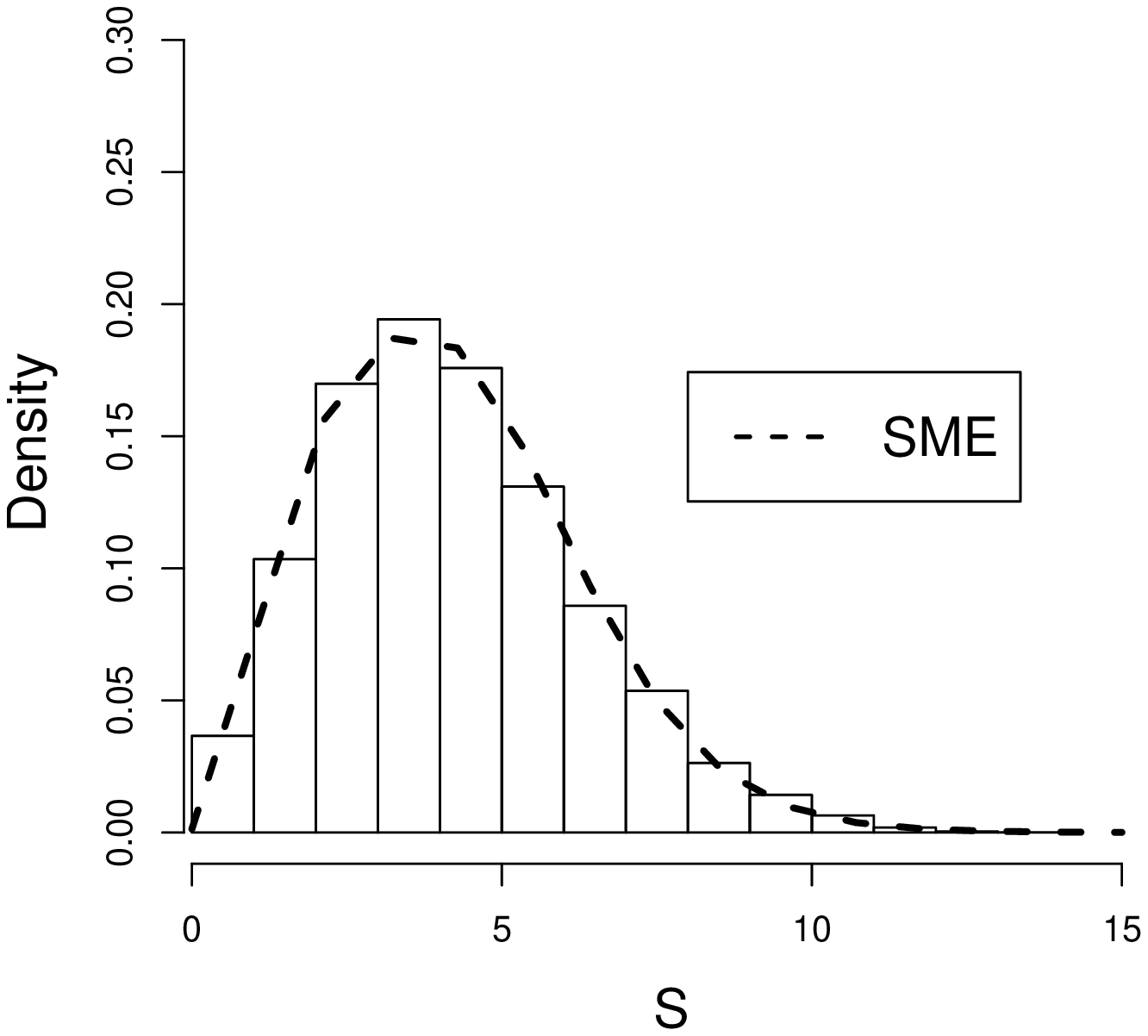}
                \caption{Case (3): \\ $\ell=4$, $\mu=0$, $\sigma=0.25$}
                \label{fig:ex5}
        \end{subfigure}%

        \begin{subfigure}[b]{0.4\textwidth}
                \includegraphics[width=\textwidth]{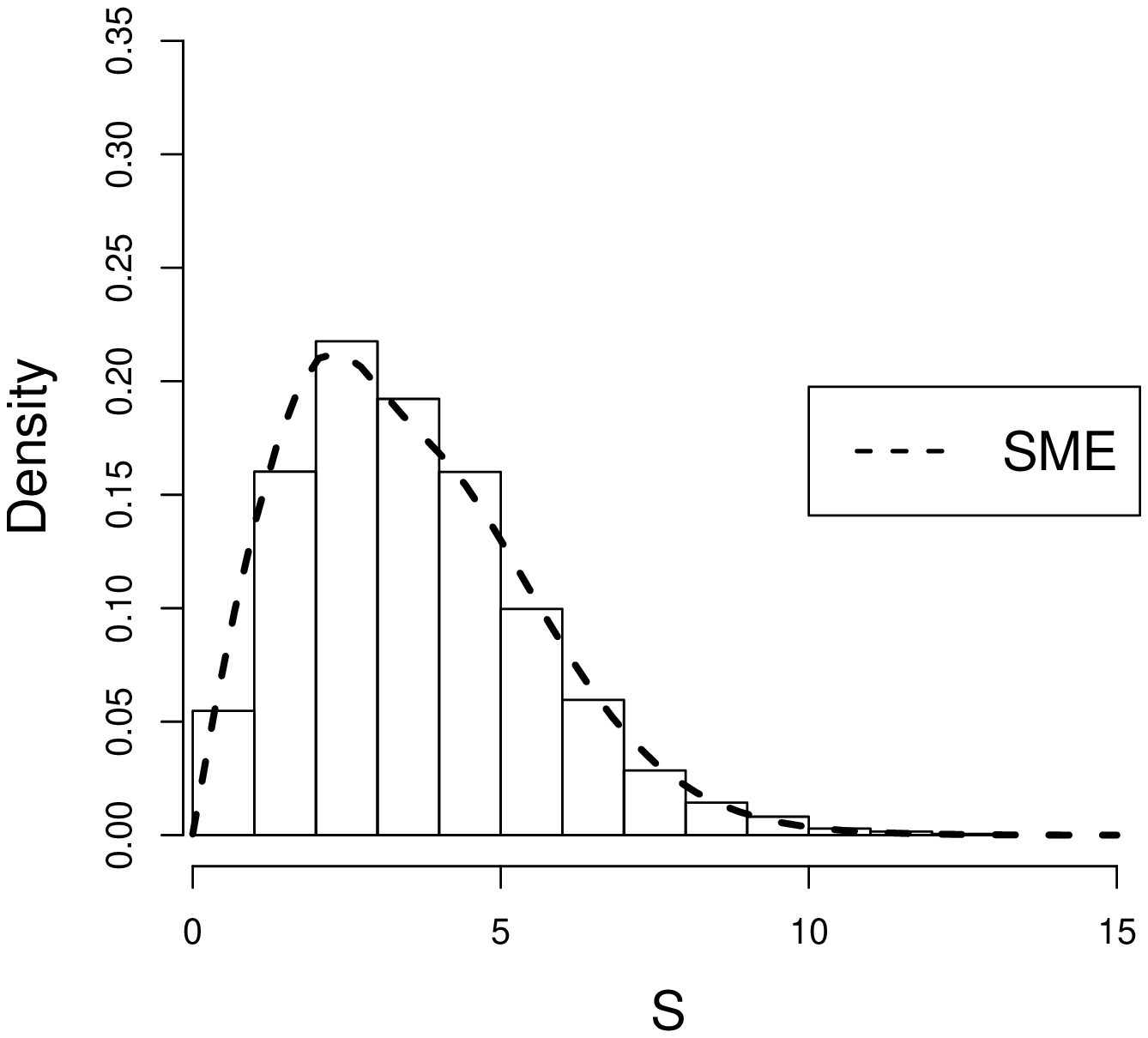}
                \caption{Case (4): \\ $\ell=3$, $\mu=0.1$, $\sigma=0.25$}
                \label{fig:ex6}
        \end{subfigure}
        \begin{subfigure}[b]{0.4\textwidth}
                %\addtocounter{subfigure}{4}
                \includegraphics[width=\textwidth]{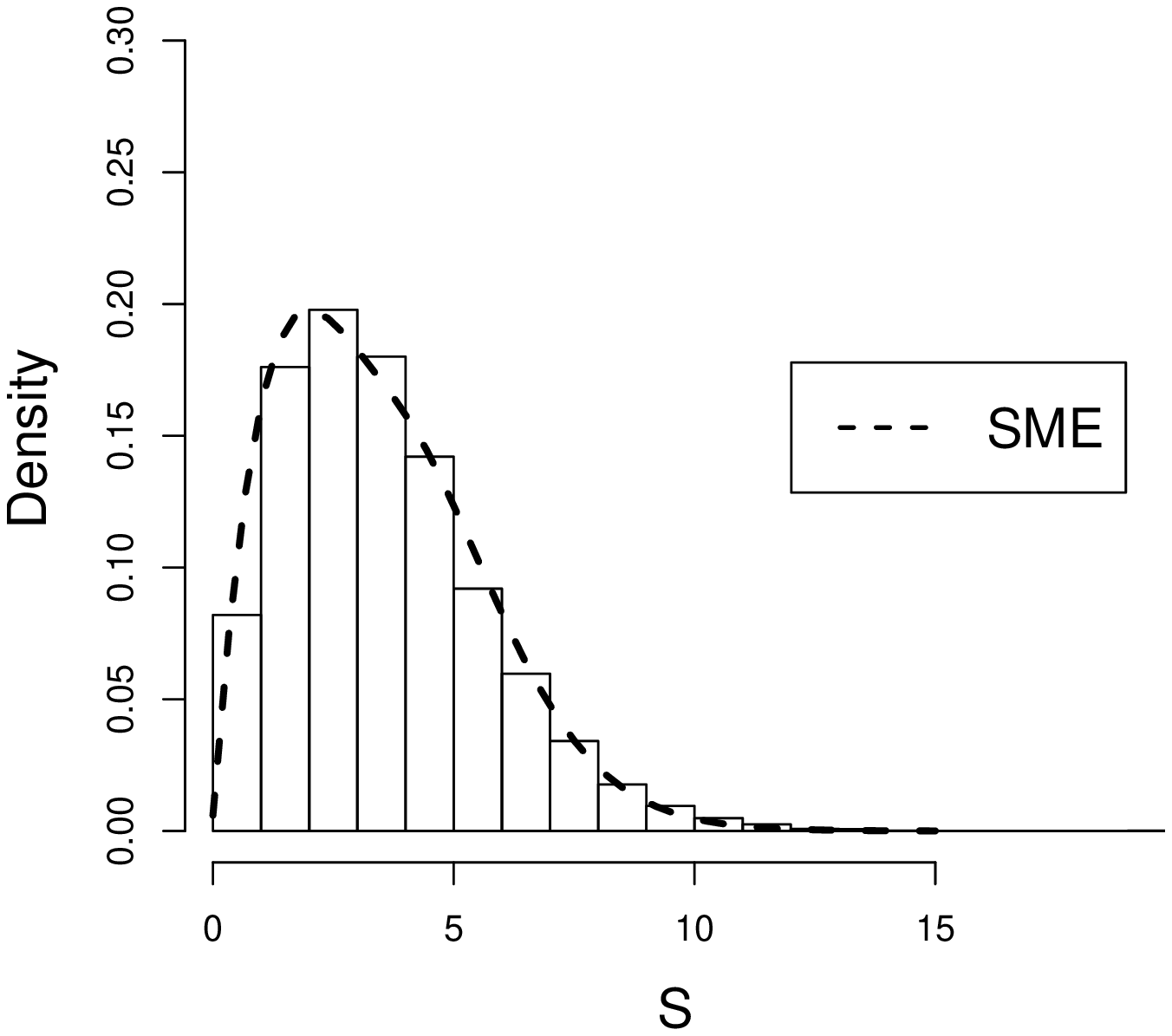}
                \caption{Case (5): \\ $\ell=3$, $\mu=0$, $\sigma=0.5$}
                \label{fig:ex8}
        \end{subfigure}
        \caption{Combined plot of densities of SME approach for compounded sums $S$, for different parameters}
        \label{SME_all_cases}
\end{figure}

 For example, the distances between the estimated and the empirical densities in the $L_1$ and $L_2$ norms, and the MAE and RMSE values, for Cases (2) to (5)  considered in Figure  (\ref{tab:SME_errors}), are listed in Table (\ref{tab:SME_errors}). They suggest that the reconstructions are reasonably good. Besides, in Table (\ref{tab:SME_errors}) we can see that in Cases (3) and (4) the values of the $L_1$, $L_2$, MAE and RMSE are smaller than those observed in Case (1), and in Cases (2) and (5) are larger.

\begin{table}[h!]
\small
\centering
\begin{tabular}{|c|c|c|c|c|}
  \hline
  Error    &  Case (2) &   Case (3) &  Case (4)  &   Case (5)                     \\ \hline
  $L_1$-norm  &  0.2649   & 0.0947     &  0.1196    & 0.1105           \\
  $L_2$-norm  &  0.2099   & 0.0399     &  0.0563    & 0.0516 \\
  MAE      &  0.0216   & 0.0038     &  0.0074    & 0.0058 \\
  RMSE     &  0.0257   & 0.0047     &  0.0094    & 0.0064 \\
  \hline
\end{tabular}
\caption{Errors SME approach, Cases (2)-(5)}
\label{tab:SME_errors}
\end{table}

In order to evaluate the ability of the SME method to perform well on unobserved data, we calculate the error and distances of the SME densities with a test data set, that is, an independent and smaller sample which comes from the same population as the observed data. The corresponding $L_1$, $L_2$, MAE and RMSE results for all the cases consider in this paper are showed in Table (\ref{tab:errors_SME_val}). They seem to show similar results to the obtained with the observed data set displayed in Tables (\ref{tab:error_case_a}) and (\ref{tab:SME_errors}). These indicate that the obtained maxentropic approximation does not suffer of overffiting and that performs well on unobserved data.

\begin{table}[h!]
\small
\centering
\begin{tabular}{|c|c|c|c|c|c|}
  \hline
  Error    & Case (1) &  Case (2) &   Case (3) &  Case (4)  &   Case (5)                     \\ \hline
  L1-norm  & 0.1223 & 0.2103    &   0.0960   &  0.1206    &  0.1471         \\
  L2-norm  & 0.0649 & 0.1847    &   0.0408   &  0.0591    &  0.0651 \\
  MAE      & 0.0109 & 0.0216    &   0.0126   &  0.0095    &  0.0120 \\
  RMSE     & 0.0147 & 0.0259    &   0.0140   &  0.0121    &  0.0171 \\
  \hline
\end{tabular}
\caption{Errors SME approach (validation set)}
\label{tab:errors_SME_val}
\end{table}

For each choice of parameters,  we applied the different statistical tests described in Section (3) and the Appendix. The results obtained are displayed in Table (\ref{critical_values1}). Three stars mean that at 1\% of significance we do not reject the null hypothesis of no differences between the empirical and maxent density, and two stars mean that we do not reject the null hypothesis at 5\% of significance. The critical values used for all the tests are indicated in the Appendix.  Most of the cases pass the tests at 5\% of significance, and the rest of the tests do it at 1\% of significance.

\begin{table}
\centering
\small
\begin{tabular}{|c|c|c|c|c|c|}

  \multicolumn{5}{c}{}\\ \hline
  \textbf{Criterion}               & Case (1)      &  Case (2)    &  Case (3)     &  Case (4)   &  Case (5)  \\ \hline
  KS test of uniformity            &  1.51***      & 1.28**   &  0.87**    &  1.17**      &  1.24**  \\
  Anderson-Darling test:           &  1.95**       & 2.44**   &  1.46**    &  1.51**      &  1.39** \\
  Cram\'{e}r- Von Mises test:                &  0.30**       & 0.44**   &  0.19**    &  0.13**      &  0.29** \\
  Berkowitz Test: :                &  5.74**       & 2.17**   &  4.43**    &  4.55**      &  1.19** \\
  Jarque-Bera Test :               &  1.34**       & 8.98***  &  2.93**    &  4.27**      &  3.67** \\
  \hline
\end{tabular}
\caption{\emph{Statistics} of SME approach (test set)}
\label{critical_values1}
\end{table}

\subsection{Numerical Reconstructions with the MEM approach}

In this section we describe the results of the implementation of the MEM approach described in Section (2.2).
Again we start by considering the compounded sum of a Poisson and simple lognormal for the individual losses, with parameters $\ell=3$, $\mu=0$ and $\sigma=0.25$ respectively. Here we use the MEM with a Poisson reference measure with parameters $\eta=2$,  and a partition in $[0,1]$ of size $M=200$, as was described in Section (2.2).

In Figure (\ref{MEM(pois)_results}) we display the density, the distribution function, the marginal calibration and reliability diagram of the MEM reconstruction and the observed data set. In panel (\ref{fig:ex4_MEM}) we show the SME and MEM reconstructions along with the histogram of the observed data, whereas in panels (\ref{fig:Fex4_MEM}), (\ref{fig:marginal_ex4_MEM}) and (\ref{fig:reliability_ex4_MEM}), we display the distribution function, the marginal calibration diagram and reliability diagram respectively, here we only show the MEM reconstruction along with the empirical distribution. In panel (\ref{fig:marginal_ex4_MEM}) we observe minor fluctuations about zero, with values not greater than 0.027 in absolute value. Such small fluctuations indicate a good fit.

\begin{figure}[h!]
        \centering
        \begin{subfigure}[b]{0.4\textwidth}
                \includegraphics[width=\textwidth]{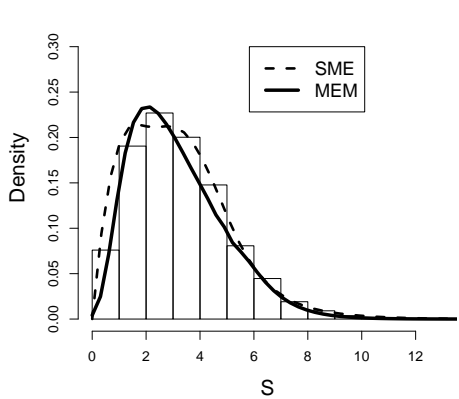}
                \caption{MEM density}
                \label{fig:ex4_MEM}
        \end{subfigure}%
        \begin{subfigure}[b]{0.4\textwidth}
                \includegraphics[width=\textwidth]{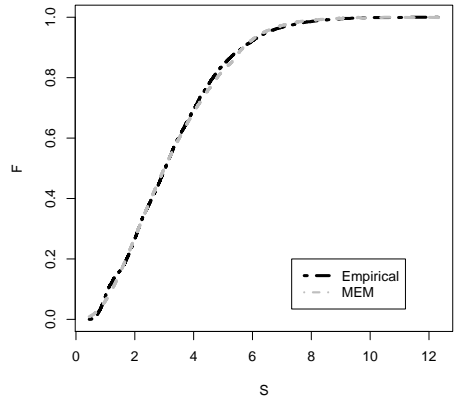}
                \caption{MEM distribution function}
                \label{fig:Fex4_MEM}
        \end{subfigure}%

        ~ %add desired spacing between images, e. g. ~, \quad, \qquad etc.
        \begin{subfigure}[b]{0.4\textwidth}
                \includegraphics[width=\textwidth]{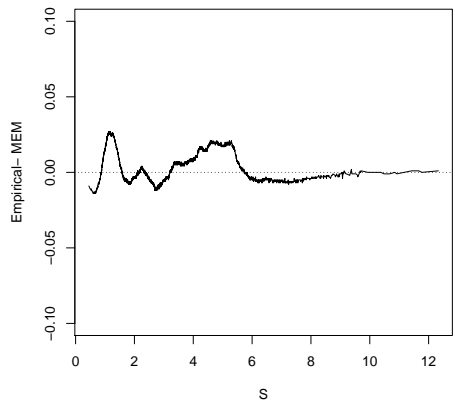}
                \caption{MEM Marginal calibration}
                \label{fig:marginal_ex4_MEM}
        \end{subfigure}
        \begin{subfigure}[b]{0.4\textwidth}
                %\addtocounter{subfigure}{4}
                \includegraphics[width=\textwidth]{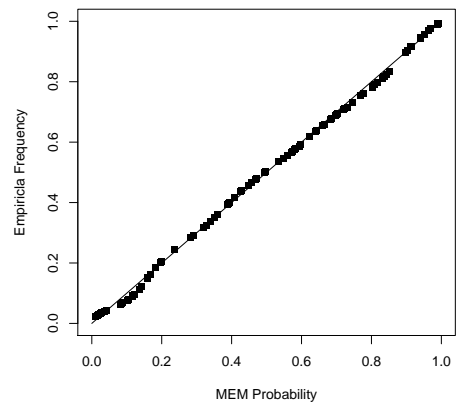}
                \caption{MEM Reliability Diagram}
                \label{fig:reliability_ex4_MEM}
        \end{subfigure}
        \caption{MEM ($\eta=2$, $M=200$) results of the case $\ell=3$, $\mu=0$, $\sigma=0.25$}
        \label{MEM(pois)_results}
\end{figure}
In Table (\ref{tab:error_case_MEM_posi}) we show the numerical distances between the reconstructions in the $L_1,$ $L_2,$ SME and  MEM distances, in order to see how different the maxentropic densities are from the observed data set. Clearly, in Table (\ref{tab:error_case_MEM_posi}) the SME reconstruction seems to be a little better than the MEM reconstruction, but even so, the MEM approximation provides us with a quite good reconstruction.

\begin{table}[h!]
\small
\centering
\begin{tabular}{|c|c|c|c|c|c|c|}
\hline
   Approach    & L1      &   L2       & MAE      & RMSE                    \\ \hline %&   L1'      &      L2'
     SME       & 0.1225  & 0.0598    & 0.0071   & 0.0089       \\                       %&   0.0102   &   0.0172
   MEM & 0.1279  & 0.0609   & 0.0086   & 0.0109       \\                         % &   0.0106   &   0.0176
  \hline
\end{tabular}
\caption{Errors of SME and MEM reconstructions \\ Case (1): $\ell=3$, $\mu=0$, $\sigma=0.25$}
\label{tab:error_case_MEM_posi}
\end{table}

In Figure (\ref{PIT(ex4_MEM_exp)}) we display the PIT transformation and correlograms of different powers for the Case (1). As we can see, the PIT histogram seems to be uniform and the correlation plot does not show any sign of dependence.

\begin{figure}[h!]
  \centering
  \includegraphics[width=9cm,height=7cm]{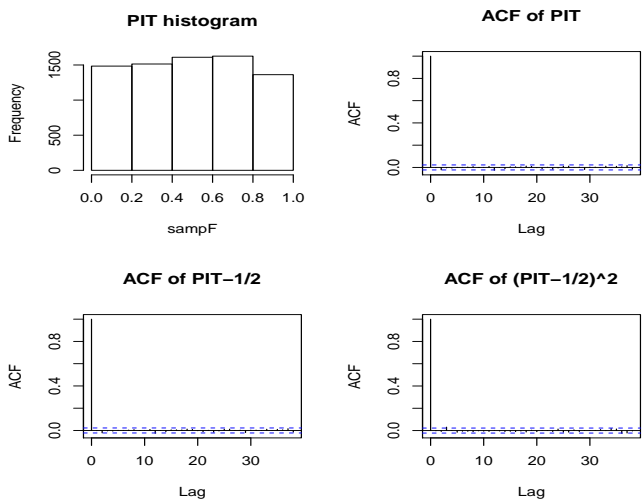}
  \caption{Probability integral transform (PIT) histogram and sample autocorrelation functions
for the MEM approach. Case (1) $\ell=3$, $\mu=0$, $\sigma=0.25$.}
  \label{PIT(ex4_MEM_exp)}
\end{figure}

In Figure (\ref{MEM_Poisson}) we display the results of MEM and SME approaches, along with the empirical histogram of the compounded sum for different values of the parameters $\ell$, $\mu$ \& $\sigma$, in order to see the differences between the reconstructions and the observed data. The reconstructions seems to fit the histograms, it is clear that there is little difference between the reconstruction. We point out that the results of the MEM method are very sensitive to the interpolation scheme used to display the results and to the size of the partition employed, which makes the tails of the distribution more difficult to adjust.

\begin{figure}[h!]
        \centering
        \begin{subfigure}[b]{0.4\textwidth}
                \includegraphics[width=\textwidth]{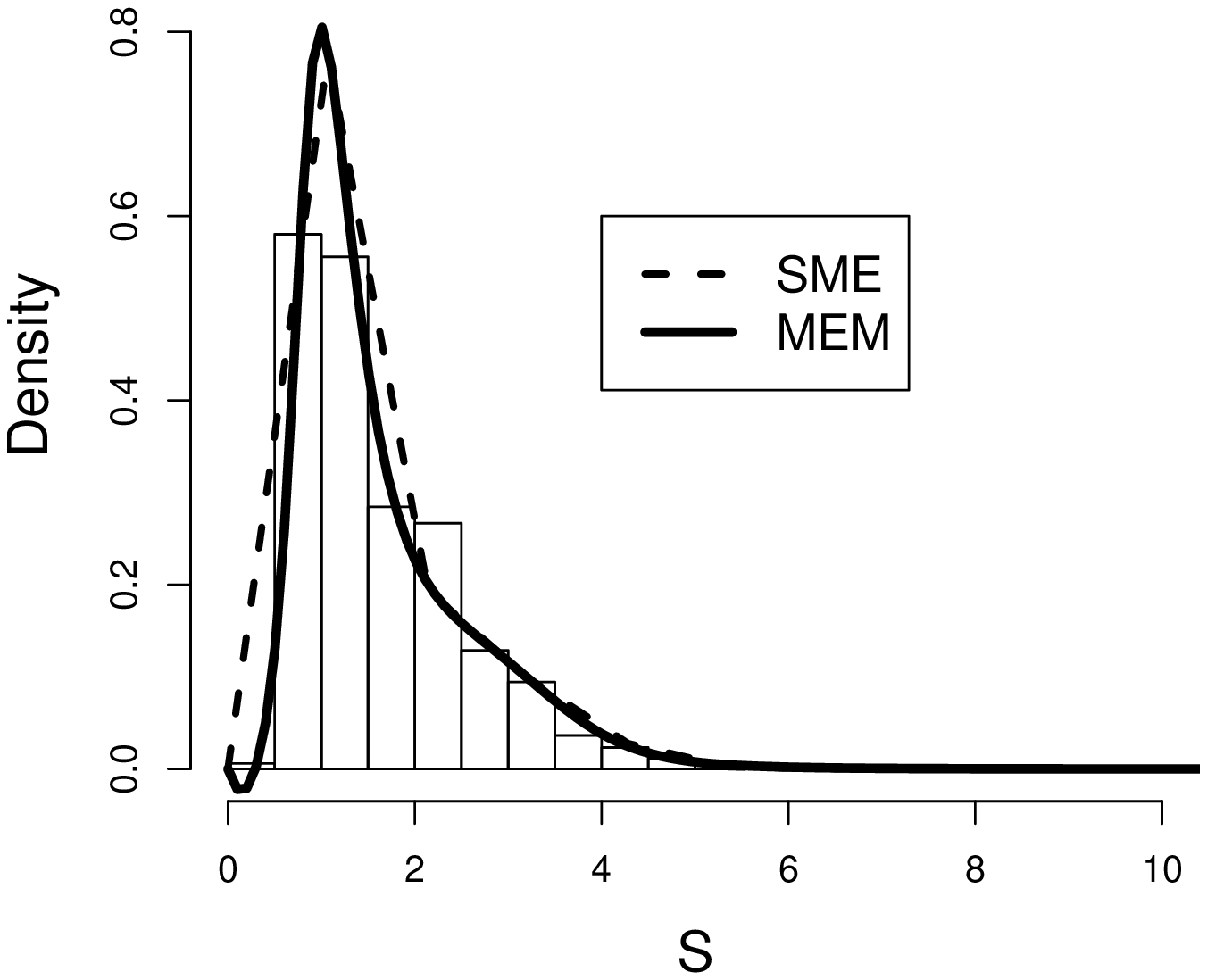}
                \caption{Case (2): \\ $\ell=1$, $\mu=0$, $\sigma=0.25$}
                \label{fig:den2}
        \end{subfigure}%
        \begin{subfigure}[b]{0.4\textwidth}
                \includegraphics[width=\textwidth]{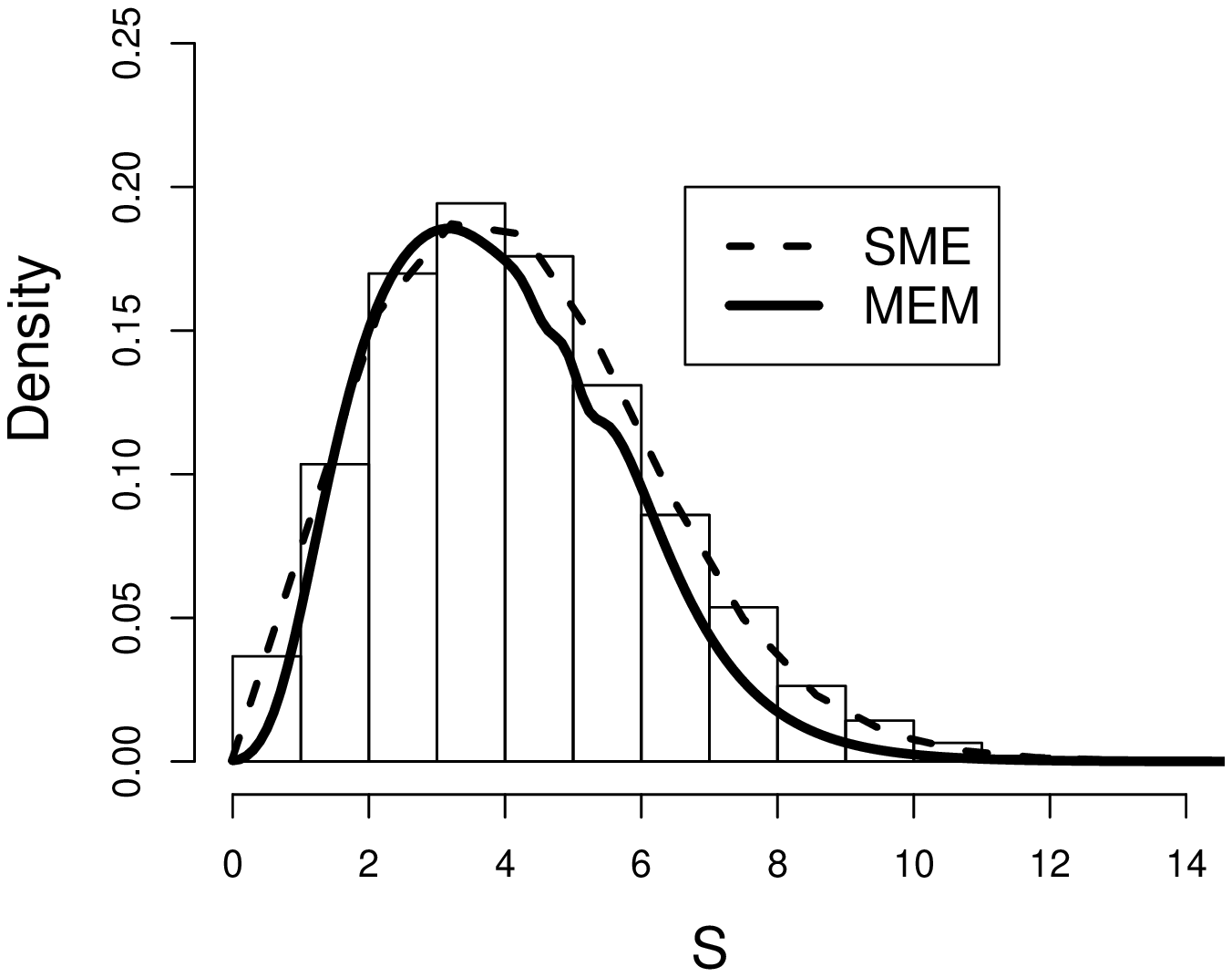}
                \caption{Case (3): \\ $\ell=4$, $\mu=0$, $\sigma=0.25$}
                \label{fig:den10}
        \end{subfigure}%

        ~ %add desired spacing between images, e. g. ~, \quad, \qquad etc.
        \begin{subfigure}[b]{0.4\textwidth}
                \includegraphics[width=\textwidth]{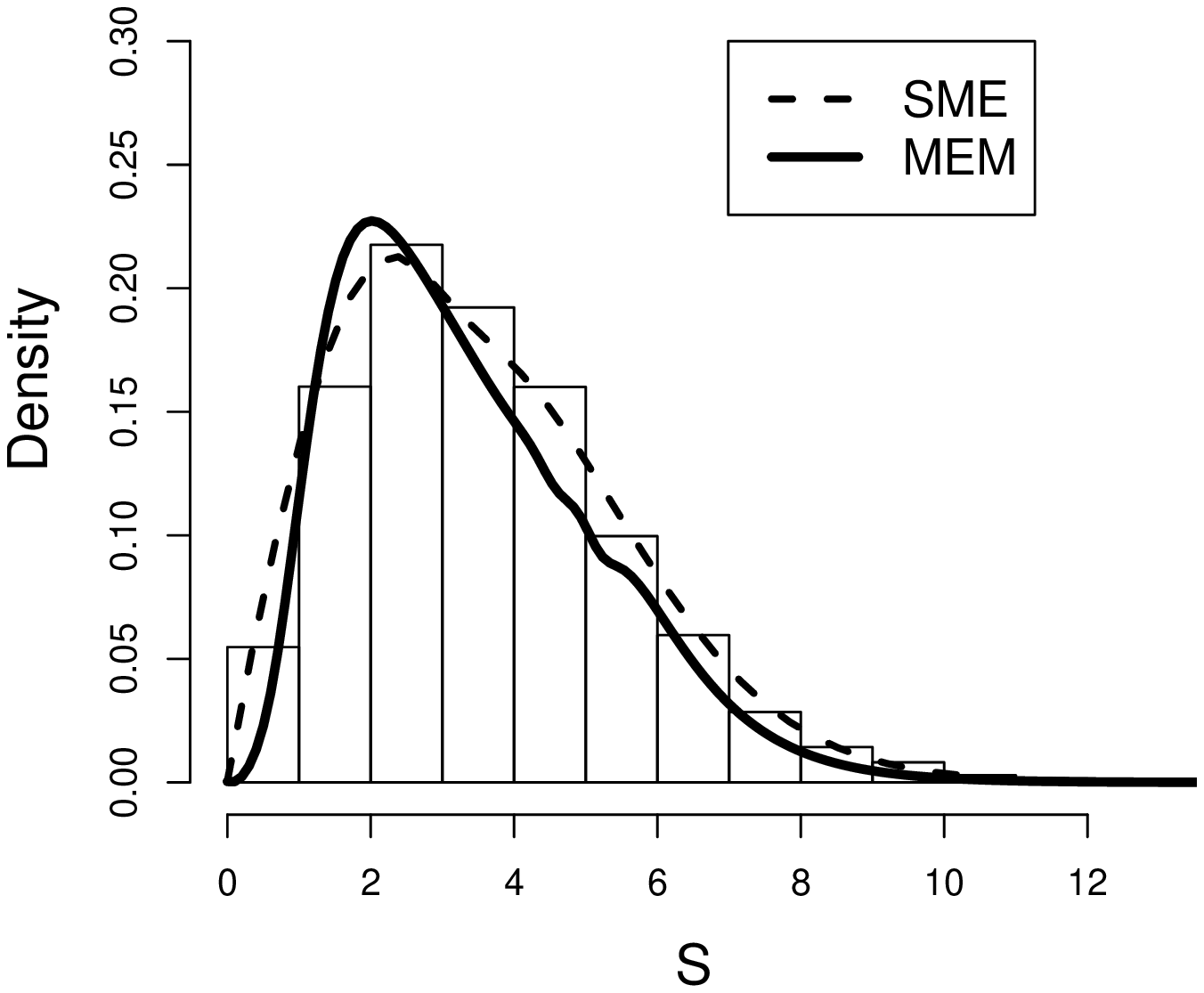}
                \caption{Case (4): \\ $\ell=3$, $\mu=0.1$, $\sigma=0.25$}
                \label{fig:den4}
        \end{subfigure}
        \begin{subfigure}[b]{0.4\textwidth}
                %\addtocounter{subfigure}{4}
                \includegraphics[width=\textwidth]{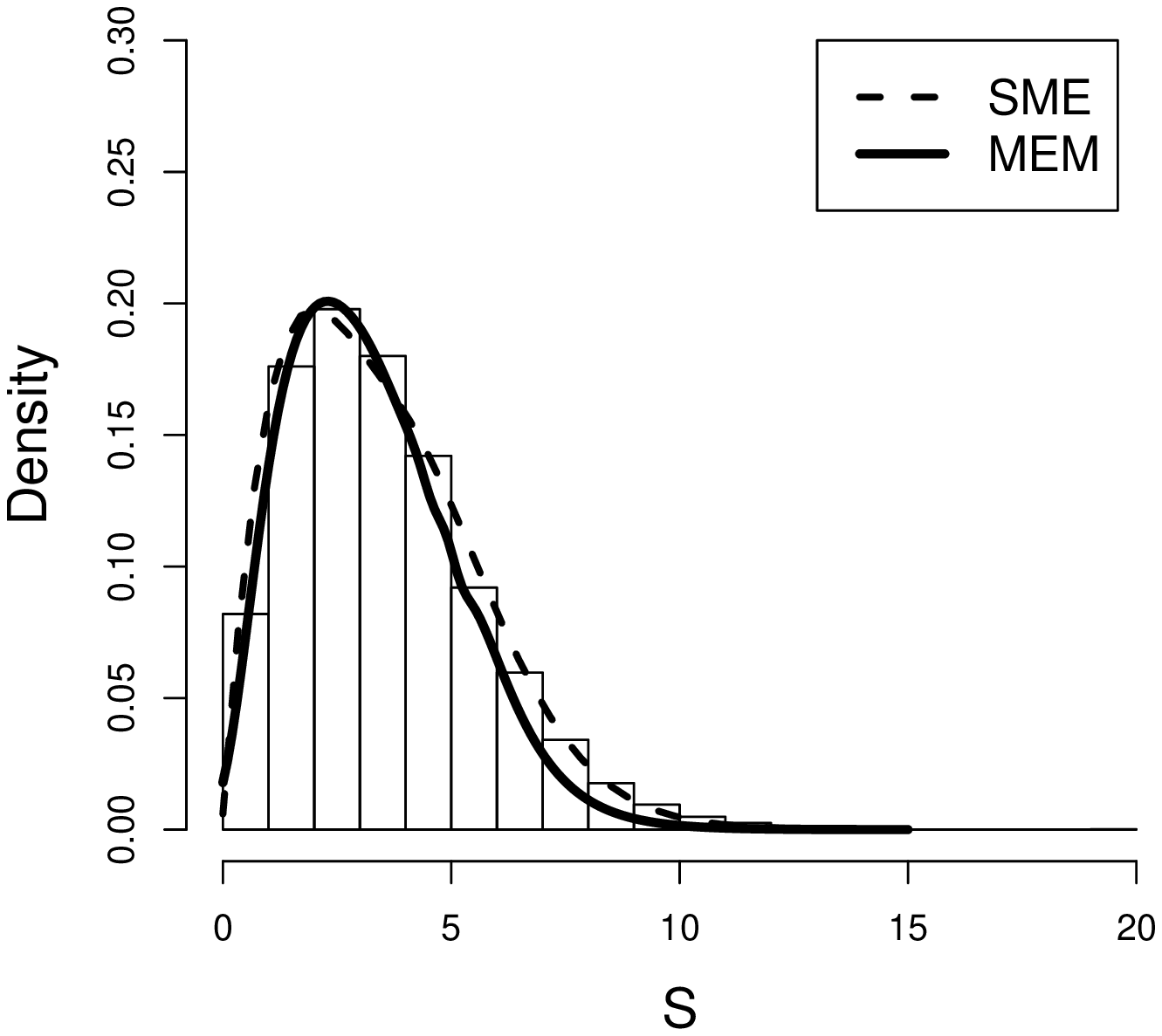}
                \caption{Case (5): \\ $\ell=3$,  $\mu=0$, $\sigma=0.5$}
                \label{fig:den6}
        \end{subfigure}
        \caption{Combined plot of SME \& MEM densities}
        \label{MEM_Poisson}
\end{figure}

In Table (\ref{tab:errors_MEM_POISSON}) we list the $L_1$ and $L_2$ distances between reconstructed and empirical densities, along with the  MAE and RMSE distances between the MEM-distribution function and the empirical distribution function. These errors are in most of the cases a little larger than those obtained for the SME reconstructions. An exception occurs for the Case (2) where the MEM reconstruction is better than the density obtained with the SME.

\begin{table}[h]
\small
\centering
\begin{tabular}{|c|cc|cc|cc|cc|}
  \hline
   \multirow{2}{*}{\textbf{ERROR}} & \multicolumn{2}{c|}{\textbf{Case (2)}} & \multicolumn{2}{c|}{\textbf{Case (3)}} & \multicolumn{2}{c|}{\textbf{Case (4) }} &\multicolumn{2}{c|}{\textbf{Case (5) }} \\ \cline{2-9}
      & SME    & MEM      & SME &  MEM                & SME  &   MEM         &  SME         & MEM                   \\ \hline
  L1-norm  & 0.2649 & 0.2560   & 0.0947 & 0.1952      & 0.1196  &  0.1652    & 0.1105       & 0.1498 \\
  L2-norm  & 0.2099 & 0.2091   & 0.0399 & 0.0857      & 0.0563  &  0.0770    & 0.0516       & 0.0605 \\
  MAE      & 0.0216 & 0.0182   & 0.0038 & 0.0172      & 0.0074  &  0.0123    & 0.0058       & 0.0114 \\
  RMSE     & 0.0257 & 0.0221   & 0.0047 & 0.0248      & 0.0094  &  0.0145    & 0.0064       & 0.0166 \\
  \hline
\end{tabular}
\caption{Errors SME and MEM  approaches, Cases (2)-(5)}
\label{tab:errors_MEM_POISSON}
\end{table}

Using the test set we can observe how does the MEM reconstruction perform on unobserved data. In Table (\ref{tab:ERRORS_VAL_POISS_MEM}) we display the values of the $L_1,$  $L_2,$ MAE and RMSE measures of error.  Clearly these results are similar to those displayed in Table (\ref{tab:errors_MEM_POISSON}) proving the good performance of the obtained reconstructions.

\begin{table}[h]
\small
\centering
\begin{tabular}{|c|c|c|c|c|c|}
  \hline
  Error    & Case (1)  &  Case (2) &   Case (3) &  Case (4)  &   Case (5)                     \\ \hline
  L1-norm  & 0.1896  & 0.2057    &  0.1580    &  0.1598    &  0.1751         \\
  L2-norm  & 0.1370  & 0.1866    &  0.0781    &  0.0763    &  0.0704 \\
  MAE      & 0.0131  & 0.0186    &  0.0201    &  0.0170    &  0.0161 \\
  RMSE     & 0.0150  & 0.0225    &  0.0223    &  0.0201    &  0.0198 \\
  \hline
\end{tabular}
\caption{Errors MEM approach (test set)}
\label{tab:ERRORS_VAL_POISS_MEM}
\end{table}

In Table (\ref{cv_val_pois}) we display the results of statistical tests applied to the test set. As in Section (4.1) those market with asterisks do no reject the null hypothesis of equality between distributions. As we can appreciate, most of the tests seems to validate our reconstruction at 5\% of significance.

\begin{table}
\centering
\tiny
\begin{tabular}{|c|c|c|c|c|c|}
  \hline
  \multicolumn{5}{c}{\textbf{Critical Values}}\\ \hline \hline
  \textbf{Criterion}               & Case (1)      &  Case (2) &  Case (3)  &  Case (4)  &  Case (5)  \\ \hline
  KS test of uniformity            & 1.07**        & 1.11**    & 1.22**     &  1.53***      &  1.23**  \\
  Anderson-Darling test:           & 2.28**        & 1.68**    & 3.71***    &  3.31***      &  2.99*** \\
  Cram\'{e}r-v. Mises test:                & 0.30**        & 0.31**    & 0.41**     &  0.25**       &  0.39** \\
  Berkowitz Test:                  & 7.74**        & 5.27**    & 10.04***   &  6.94**       &  1.54** \\
  Jarque-Bera Test:                & 10.25         & 7.78***   & 3.75**     &  2.28**       &  9.20*** \\
  \hline
\end{tabular}
\caption{\emph{Critical Values} of MEM approach (test set)}
\label{cv_val_pois}
\end{table}

\section{Risk measures}

 In this Section we present the computation of the $VaR$ and $TVaR$ of the total loss $S$ corresponding to the parameters $\ell=3$, $\mu=0$ and $\sigma=0.25$ (i.e., Case (1)),  at various confidence levels. This results also serves to test the  potential of the SME and MEM approaches. In this example we compare the quantiles of the reconstruction against the sample quantiles.

 For the calculation of $VaR$ and $TVaR$ at the confidence level $\gamma$ of the SME and MEM reconstructions, we use the theoretical definition of VaR and TVaR  simplified in the  following lemma taken from Rockafellar and Uryasev (2000).

\begin{lemma}
 The function
$$a\rightarrow U(a) = a+\frac{1}{1-\gamma} \int\limits_{a}^\infty (t-a) f^*_S(t)dt $$
\noindent defined on $(0,\infty)$ is convex in $a$, achieves its minimum at $VaR_{\gamma}$ and its minimum value
is $TVaR_{\gamma}(S)=E^*[S|S>VaR_{\gamma}]$. Above,  $f^*_S$ denotes the maxentropic density.
\end{lemma}

In Tables (\ref{VaR}) and (\ref{TVaR}) we display the resulting computations for a collection of $\gamma$'s. The lasts columns are corresponding to $VaR$ and $TVaR$ of the simulated sample (observed data set) along with their confidence levels at 95\%. Additionally we include the absolute difference between the VaR and TVaR of the observed data with those from the VaR and TVaR obtained from the reconstructions.

In order to calculate the empirical VaR and TVaR we consider $S>0$ ordered in increasing
size ($s_1 \leq s_2 \leq s_n$), and then estimated
$VaR$ as $\widehat{VaR}_{\gamma}(S)\approx x([N(\gamma)])$, where $[a]$ denotes the integer part of the real number $a.$ The estimate of the TVaR is obtained from the same ordered list of values as $$\widehat{TVaR}_{\gamma}= \frac{1}{N-[N{\gamma}]+1} \sum\limits_{j=[N(\gamma)]}^N s_j.$$ The confidence intervals for the VaR and TVaR were calculated by resampling without replacement using subsamples of size 90\% of the total data size.

 \begin{table}[h!]
 \centering
 %\small
\begin{tabular}{|c|ccc|cc|cc|}
\hline
\multicolumn{8}{|c|}{VaR}\\ \hline \hline
   & \multicolumn{3}{c}{Approaches} \quad \quad \quad  & \multicolumn{2}{|c}{Errors} & \multicolumn{2}{|c|}{Confidence Interval}\\  \cline{2-8}
$\gamma$  & SME    & MEM             & Empirical    & SME error & MEM error   &  $VaR_{inf}$ &   $VaR_{sup}$     \\ \hline
  0.900  & 5.657* &   5.657*           & 5.672        & 0.015     & 0.015       & 5.587 & 5.763  \\
  0.910  & 5.798* &   5.818*           & 5.809        & 0.011     & 0.009       & 5.725 & 5.920  \\
  0.920  & 5.939* &   5.980*           & 5.968        & 0.029     & 0.012       & 5.872 & 6.055  \\
  0.930  & 6.081* &   6.141*           & 6.118        & 0.037     & 0.023       & 6.021 & 6.227 \\
  0.940  & 6.222* &   6.303*           & 6.299        & 0.077     & 0.004       & 6.202 & 6.379  \\
  0.950  & 6.505* &   6.465*           & 6.474        & 0.031     & 0.009       & 6.377 & 6.614  \\
  0.960  & 6.788* &   6.788*           & 6.759        & 0.029     & 0.029       & 6.617 & 6.908  \\
  0.970  & 7.071* &   7.273*           & 7.122        & 0.051     & 0.151       & 6.955 & 7.334  \\
  0.980  & 7.495* &   7.919           & 7.583        & 0.088     & 0.336       & 7.428 & 7.767  \\
  0.990  & 8.485* &   8.566*           & 8.384        & 0.101     & 0.182       & 8.078 & 8.593  \\
  0.995  & 9.051* &   9.061*           & 9.016        & 0.035     & 0.045       & 8.747 & 9.210 \\
  0.999  & 9.192 &   10.182*          & 10.34        & 1.148     & 0.158       & 9.686  & 11.43  \\ \hline
\end{tabular}
\caption{Comparison of VaR for the SME and MEM reconstructions, \\ Case (1): $\ell=3$, $\mu=0$, $\sigma=0.25$}
\label{VaR}
 \end{table}

 \begin{table}[h!]
 \centering
 \small
\begin{tabular}{|c|ccc|cc|cc|}
\hline
\multicolumn{8}{|c|}{TVaR}\\ \hline \hline
   & \multicolumn{3}{c}{Approaches} \quad \quad \quad  & \multicolumn{2}{|c}{Errors} & \multicolumn{2}{|c|}{Confidence Interval}\\ \cline{2-8}
$\gamma$ & SME      & MEM       & Empirical   & SME error & MEM error & $TVaR_{inf}$ &   $TVaR_{sup}$     \\ \hline
  0.900  & 6.817*   & 6.833*     & 6.839       & 0.022     & 0.006     & 6.721  & 6.948 \\
  0.910  & 6.930*   & 6.967*     & 6.961       & 0.031     & 0.006     & 6.846  & 7.079 \\
  0.920  & 7.055*   & 7.041*     & 7.095       & 0.040     & 0.054     & 6.974  & 7.232\\
  0.930  & 7.192*   & 7.121*     & 7.245       & 0.053     & 0.124     &  7.109 & 7.389 \\
  0.940  & 7.344*   & 7.303*     & 7.417       & 0.073     & 0.114     &  7.276 & 7.562\\
  0.950  & 7.513*   & 7.528*     & 7.622       & 0.109     & 0.094     & 7.459  & 7.781 \\
  0.960  & 7.935*   & 7.814*     & 7.874       & 0.061     & 0.060     & 7.700  & 8.045 \\
  0.970  & 8.205*   & 8.220*     & 8.188       & 0.017     & 0.032     & 7.989  & 8.373 \\
  0.980  & 8.533*   & 8.516*     & 8.601       & 0.068     & 0.085     & 8.405  & 8.817 \\
  0.990  & 8.959*   & 8.912*     & 9.262       & 0.303     & 0.350     & 8.968  & 9.555 \\
  0.995  & 9.556*   & 9.606*     & 9.843       & 0.287     & 0.237     & 9.465  & 10.222 \\
  0.999  & 10.543*  & 11.215*    & 11.167      & 0.624     & 0.048     & 10.341 & 11.848 \\ \hline
\end{tabular}
\caption{Comparison of TVaR for the SEM and MEM reconstructions, \\ Case (1): $\ell=3$, $\mu=0$, $\sigma=0.25$}
\label{TVaR}
 \end{table}

 In Tables (\ref{VaR}) and (\ref{TVaR}) the asterisks indicate that the calculated VaR and TVaR for the reconstructions belongs to the empirical confidence interval at the $95\%$ level, also shown in each table. The rather good agreement of the results displayed in these tables are a further indication of the quality of the maxentropic methods.

\section{Decompounding}

It may not be always possible to observe frequency and severity separately. That is, even though the frequency of events is recorded, only the total or aggregated loss data is available. Nevertheless, the risk analyst may want to know the distribution of individual losses, because it is at that level where loss prevention or mitigation is applied. The maxentropic approaches that we have developed here allow us to determine the distribution of the individual loss as well.

In our case, we know how to compute or we can estimate numerically, the Laplace transform $\psi(\alpha)$ of the total losses $S,$ and we also have an analytic expression for the generating function $G(z)$ of the frequency of the events $N.$ From these ingredients, we can obtain the Laplace transform $\phi(\alpha)$ of the individual losses, which we can use as starting point to determine the probability distribution of individual losses. The relationship between the various Laplace transforms is contained in equation (\ref{eq3}), from which we obtain
\begin{center}
$\phi(\alpha_k)=\frac{1}{l} ln (\psi(\alpha_k)) +1$,
\end{center}
\noindent where, to use what we have developed above, we may write
$$\psi(\alpha_k)=e^{-\ell}+(1-e^{\ell})\int\limits_{0}^1 y^{\alpha_k} f^*_Y(y)dy,$$
for the Laplace transform of the aggregate losses, $\ell$ being the  parameter of the observed Poisson frequency, which is known in our case, and $f_Y^*$ is the maxentropic probability density of the total losses, which we have already determined.

To exemplify our procedure, we shall use Case (1), characterized by the following parameters: $\ell=3$, $\mu=0$ and $\sigma=0.25.$ Having  determined numerically $f^*_S,$ and computed $\phi(\alpha)$ as mentioned above, we can apply  the SME and MEM procedures to obtain the probability density of individual losses. The resulting individual densities are shown in Figure (\ref{Individuals}). The comparison with the known probability density that was used to generate the total losses is another consistency test of the procedure. The true distributions of individual losses is a lognormal density with parameters $\mu=0$, $\sigma=0.25$, and is included in Figure (\ref{Individuals}).

\begin{figure}[h!]
  \centering
  \includegraphics[width=9cm,height=7cm]{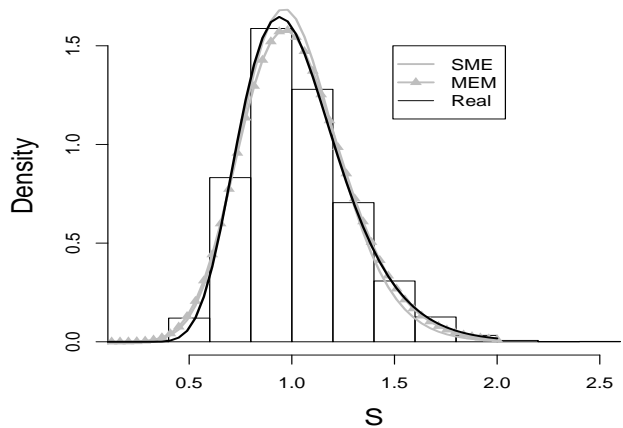}
  \caption{Density of the individual losses obtained by SME \& MEM approaches\\ Case (1): $\ell=3$, $\mu=0$, $\sigma=0.25$}
  \label{Individuals}
\end{figure}

 In Tables (\ref{X1}) and (\ref{X2}) we present the results of several measures of quality of reconstruction. Here the SME and MEM reconstructions are compared with the true lognormal density and with the observed (simulated) data. We include the comparison between the reconstructed and true density to the data obtained by simulation, but keep in mind that such data might not be available. The true density is to be used as a benchmark to test the quality of the maxentropic procedures. Note that the best results are obtained for the MEM-reconstruction, whose errors are smaller, namely, $0.0087$ and $0.0105$ respectively, as measured by the MAE and RMSE distances to the empirical data.

 The $L_1$ and $L_2$ distances between the reconstructions by both procedures and the histogram are similar to those between the true distribution and the histogram. Notice that in the left column of the second table are equal because they measure distance between the true density and the histogram.  The distances between the maxentropic density and the histogram, measured by the  $L_1$ and the $L_2$ are $0.16$ and $0.18$ respectively. Notice that they are similar to the same distances between the true density and the histogram. These good results play the role of a check up test for the SME and MEM procedures applied throughout this paper.

\begin{table}[h!]
\centering
\small
\begin{tabular}{|c|cc|cc|cc|}
\hline
\multirow{2}{*}{Approach} & \multicolumn{2}{c|}{Hist. vs. True density} & \multicolumn{2}{c|}{Hist. vs. Maxent} & \multicolumn{2}{c|}{Real density vs. Maxent}
\\ \cline{2-7}
                         & MAE      & RMSE   & MAE      & RMSE     & MAE     & RMSE      \\ \hline
SME                      & 0.0042  & 0.0047 &  0.0127   &  0.0257  & 0.0143  &  0.0232    \\
MEM                      & 0.0042  & 0.0047 &  0.0087   &  0.0105  & 0.0096  &  0.0113    \\
 \hline  \cline{1-3}
\end{tabular}
\caption{MAE and RMSE values of the individual losses calculated by SME \& MEM approaches}
\label{X1}
\end{table}

\begin{table}[h]
\centering
\begin{tabular}{|c|cc|cc|cc|}
\hline
\multirow{2}{*}{Approach} & \multicolumn{2}{c|}{Hist. vs. True density} & \multicolumn{2}{c|}{Hist. vs. Maxent} & \multicolumn{2}{c|}{Real density vs. Maxent}  \\ \cline{2-7}
                          & L1-norm      & L2-norm   & L1-norm      & L2-norm   & L1-norm      & L2-norm   \\ \hline
SME                       & 0.1640       & 0.1865     & 0.1886       & 0.1992    & 0.0679       & 0.0574  \\
MEM                      & 0.1640       & 0.1865     & 0.1659       & 0.1818    & 0.0621       & 0.0624        \\
\hline  \cline{1-3}
\end{tabular}
\caption{$L_1$ and $L_2$ distances of the individual losses calculated by SME \& MEM approaches}
\label{X2}
\end{table}

\section{Final comments and Conclusions}
We examined the performance of two maxentropic methods for the reconstruction of the probability density of total losses in a standard actuarial model of interest to the insurance and banking industries. The two methods are based on the maximization of entropy functionals defined on an appropriate space of probabilities. For both procedures the input are the values of the Laplace transform of the total loss distribution, which in our case has been determined numerically.

 One important reason for using the maxentropic methods, is that they provide good reconstructions based on a very small amount of information. In our case this means the knowledge of the Laplace transform at $8$ real values of its transform's parameter. Besides providing quite reasonable representations of the density of total loss, the maxentropic methods admit extension that can be used to handle measurement error or scarce data situations.

For the example that we analyzed from Section (4) on, is one for which an analytical solution is not known, and the standard numerical procedures to determine it, are harder to implement that those we developed here.

To determine the quality of our method, we carried out a variety of statistical tests, including a consistency test consisting of an application of a decompounding procedure to determine the probability density of the individual losses when only the total loss is recorded and a model is available for the frequency of events. The statistical consistency tests are passed quite satisfactorily.

Another reason for using these approaches is that the maxentropic procedure can be extended to the case in which there are errors in the data. Such errors in our case are associated to the size of the data, which for example, in the operational risk losses is a frequent situation. When the amount of data is small, there is a statistical indeterminacy in the values of the moments. Thus having a method that takes this fact into account is certainly of value.

Besides the potential applicability of the method for generic statistical analysis, our interest lies in exploring further the applicability of the maxentropic methodology to problems in operational risk and insurance, specially for data that could have problems of multi-modality, heavy tails, or may include extreme events or economic shocks, for which the estimation of a probability density may be extremely difficult to compute.

\section{Appendix}

\textbf{GOODNESS-OF-FIT TESTS}

There are a few tests that help us to determine whether the reconstructed density is appropriate or not. Here we recall briefly the basic facts about the tests that we mentioned above.

\subsection{Kolmogorov-Smirnov test}

The Kolmogorov-Smirnov test is a test of uniformity which verifies the differences of fit between the empirical distribution function (EDF) and the estimated (reconstructed) distribution function,
using the largest absolute observed distance between them,
$$D_n=\underset{s}{sup} |F_n(s_j)-F^*_S(s_j)| $$
\noindent where $n$ is the number of data points; $\{s_j| j=1,...,n\}$ are the sample data points of the total losses $S$; $F_n(\cdot)$ is the (cumulative) empirical distribution function; and $F^*_S(\cdot)$ is the maxentropic (cumulative) distribution function (Cruz, 2002).

The null hypothesis $H_o$ of no difference between distributions, has to be rejected at the chosen significance level $\alpha$ of 0.1, 0.05 or 0.01 whenever $\sqrt{n} D_n> \sqrt{n} d_{\alpha,n}$ or if $p-value < \alpha$, where the critical value $d_{\alpha,n}$ and the p-value are calculated from the distribution of the K-S statistic when the null hypothesis is true. This is not a straightforward distribution, but could be obtained asymptotically or by simulation. In summary, the null hypothesis is rejected if the statistic $\sqrt{n} D_n$ is greater than 1.22, 1.36, 1.63 at the 90\%, 95\% and 99\% confidence levels respectively.

A problem with this test is that the KS statistic depends on the maximum difference, without considering the whole estimated distribution. This is important  when the differences between distributions are suspected to occur only at the upper or lower end of their range (Cruz, 2002). This may be particularly  problematic in small samples. Besides, little is known about the impact of the departures from independence on $D_n,$ i.e., if we are not sure about the independence of the sample, we would not be sure what is the meaning of the results of the test.  There exist other EDF tests, which in most situations are more effective than the simple Kolmogorov–Smirnov test. Further details about the K-S test can be seen in Stephens (1974).

\subsection{Anderson-Darling test}
This is a more sophisticated version of the KS approach, and it is based on the quadratic difference between $F_n(s)$ and $F^*_S(s)$.
Here the AD statistic is computed as follows:
$$A_n^2= n \int \limits_{-\infty}^\infty |F_n(s)-F^*_S(s)|^2\Psi(s)f(s)ds,$$
\noindent where $\Psi(s)=\frac{1}{F^*_S(s)(1-F^*_S(s))}$  is a weight function; $n$ is the number of data points, and $\{s_j| j=1,...,n\}$ the observed (simulated in our case) total loss $S$ sample, $F_n(\cdot)$ is the empirical (cumulative) distribution function and $F^*_S(\cdot)$ is the (cumulative) maxentropic distribution function.

When $\Psi(s)=1$, the Anderson-Darling (AD) statistic reduces to the statistic which its today known as the Cram\'{e}r-von Mises statistic. The test make maximum use of the observed data by integrating the vertical distances over all values of the sample $S$, increasing its power balancing the variance
between distributions through the use of $\Psi(\cdot)$. The test of Anderson emphasizes more the tails of the distribution than the KS-test does, (Cruz, 2002).

The test statistic may be assessed against critical values in order to reject or not the $H_o$ of uniformity. The null hypothesis rejects if $A_n^2$ is greater
than a critical value of 2.492 and 3.857 at the 95\% and 99\% confidence levels respectively. For the case of Cram\'{e}r-von Mises test, we reject the null hypothesis when the statistic is greater than 0.461 and  0.743 at the 95\% and 99\% confidence level respectively.

The Anderson-Darling (AD) statistic behaves similarly to the Cram\'{e}r-von Mises statistic, but is more powerful to test whether $F^*_S(s)$ departs from the true distribution in the tails, especially when there appear to be many outlying values. For goodness-of-fit testing, departure in the tails is often important to detect, and $A_n^2$ is the recommended statistic (Marsaglia et. al., 2004).

\subsection{Berkowitz back test}
Berkowitz (2001) proposed the transformation $z_n=\Phi^{-1}(\int^{s_n}_{-\infty}f^*_S(s)ds)=\Phi^{-1}(F(s_n)) $, to make the data i.i.d standard normal under the null hypothesis. This allows to make use of powerful battery of available normality tests, instead of relying on uniformity tests. Besides that, Berkowitz back test provides a joint test of normality and independence.

The procedure consists of testing the null hypothesis of $\rho=\mu=0$, $\sigma=1$, against a first-order autoregressive alternative ($z_t-\mu=\rho(z_{t-1}-\mu)+\varepsilon_t$) with mean and variance possibly different from (0,1). The LR test can be formulated as
\begin{eqnarray}  \label{equation2} %\nonumber
LR_3=-2(L(0,1,0)-L(\hat{\mu},\hat{\sigma}^2,\hat{\rho}))
\end{eqnarray}
\noindent where $L(\hat{\mu},\hat{\sigma}^2,\hat{\rho})$ is the likelihood as a function only of the unknown parameters of the model, the hats denote estimated values. The exact
function associated with the first order autoregressive alternative is reproduced here by convenience.

\begin{eqnarray} \nonumber
L(\mu, \sigma^2, \rho)= -\frac{1}{2}log(2\pi)-\frac{1}{2}log[\sigma^2/(1-\rho^2)]-\frac{(z_1-\mu/(1-\rho))^2}{2\sigma^2/(1-\rho^2)} \\
-\frac{T-1}{2}log(2\pi)-\frac{T-1}{2}log(\sigma^2)-\sum^T_{t=2}\left(\frac{(z_n-\mu-\rho z_{n-1})^2}{2 \sigma^2}\right)
\end{eqnarray}

\noindent where $\sigma^2=VAR(\varepsilon_t)$. Under the null hypothesis, the test statistic is distributed as a $\chi^2(3)$. This means that we reject the null hypothesis when the statistic is greater than 7.815 and 11.34 at the 95\% and 99\% confidence levels, respectively.

It is usually recommended to supplement the Berkowitz test with at least one additional test for normality, for example Jarque-Bera test. This extra test ensures that we test for the predicted normal distribution, and not just for the predicted values of the parameters  $\rho$, $\mu$ and $\sigma$.

\subsection{Jarque-Bera test}

 %$z_t=\Phi^{-1}(\int^{x_t}_{-\infty}f^*(u)du)=\Phi^{-1}(F(x_t)) \sim iid \quad N(0,1)$
The standard Jarque-Bera-test is a test of normality, whose statistic JB is defined by
$$JB=\left(\frac{n}{6}\right)\left(SW^2+\frac{(K-3)^2}{4}\right), $$
\noindent where the sample skewness is SW=$\hat{\mu}_3/\hat{\mu}_2^{3/2}$  and the sample
kurtosis is $K=\hat{\mu}_4 / \hat{\mu}_2^{2}$, for $\{\hat{\mu}_j: j=2,3,4\}$ which are the second, third and fourth central moments respectively, estimated by $\hat{\mu}_j=(1/n)\sum(s_i-\bar{s})^j$, $j=2,3,4$, where  $n$ is the sample size and $\{s_i| i=1,...,n\}$ are sample data points of the total losses $S$.

JB is asymptotically chi-squared distributed with two degrees of freedom because JB is just the sum of squares of
two asymptotically independent standardized normals. This means that $H_o$ has to be rejected at level $\alpha$ if $JB \geq \chi^2_{1-\alpha,2}$, being $H_o$
the null hypothesis which checks if the sample follows a normal random variable with unknown mean and variance. The critical values for reject the null hypothesis are 5.991 and 9.21 at the 95\% and 99\% confidence levels, respectively.

Unfortunately, the standard JB statistic is very sensitive to extreme observations, due to the fact that the empirical moments are known to be very sensitive to outliers; and the sample variances is more affected by outliers than the mean, disturbing the estimations of the sample skewness and kurtosis. To solve the problem a robust modification of the Jarque-Bera test was proposed by Gel at. al. (2008), which utilizes the robust standard deviation (namely the average absolute deviation from the median (MAAD)) to estimate a more robust kurtosis and skewness from the sample (Gel at. al.,2008).

The robust Jarque-Bera (RJB) test statistic is defined by

$$RJB=\left(\frac{n}{C_1}\right)\left(\frac{\widehat{\mu}^3}{J^3_n}\right)^2 +\left(\frac{n}{C_2}\right)\left(\frac{\widehat{\mu}^4}{J^4_n}-3\right)^2  $$

\noindent
where the Average Absolute Deviation from the Median (MAAD) is calculated as $J_n=\frac{\sqrt{(\pi/2)}}{n} \sum\limits_{i=1}^n |s_i-Median(s_i)|$, the robust sample estimates of the skewness and kurtosis are $\frac{\widehat{\mu}^3}{J^3_n}$ and $\frac{\widehat{\mu}^4}{J^4_n}$ respectively, $C_1$ and $C_2$ are positive numbers and $\{s_i| i=1,...,n\}$ are sample data points of the total losses $S$ (Gel at. al.,2008).

As in the standard Jarque-Bera test the  robust Jarque-Bera (RJB) test statistic asymptotically follows the $\chi^2_{1-\alpha,2}$ distribution with two degrees of freedom.  Then, the null hypothesis $H_o$, has to be rejected  if $RJB \geq \chi^2_{1-\alpha,2}$, being the critical values  5.991 and 9.21 at the 95\% and 99\% confidence levels, respectively.

\subsection{Correlograms}

Tests like KS and AD does not prove independence by itself, so to asses whether the probability integral transformation (PIT) of the data, denoted by $p_t$, is i.i.d, we use a graphical tool, the correlogram, which is a tool that helps in the detection of particular dependence patterns and can provide information about the deficiencies of the density reconstructions (Diebold et. al, 1998).

As we are interested not only in linear dependence but also
in other forms of nonlinear dependence such as conditional heteroscedasticity, we examine not only the correlogram of $(p_t-\bar{p}_t)$ but also it powers, i.e. the correlograms of
$(p_t-\bar{p}_t)^2$ and $(p_t-\bar{p}_t)^3$ which are the conditional variance and conditional skewness. This is sometimes complemented with the Ljung–Box test, it tests the overall randomness based on a number of lags (Diebold et. al, 1998).

%(z - E)4, conditional skewness

\subsection{Reliability Diagram or QQ-plots}

This plot serves to determine the quality of a fit by the proximity of the fitted curve to the diagonal, the closer the better the approximation, deviations from the diagonal gives the conditional bias. Additionally, this plot could indicates problems as overfitting, when the fitted curve lies below the diagonal line and underfitting when the fitted curve lies above the line.

To obtain the reliability diagram we need to find the maxentropic (cumulative) distribution function of $F^*_S$ against the empirical (cumulative) distribution function $F_n$  If the model is good, the points will lie very close to the line that goes from 0 to 1.

\subsection{Marginal Calibration Plot}

Another useful and similar tool is the marginal calibration plot, which is based in the idea that a system is marginally calibrated if its estimations and observations
have the same (or nearly the same) marginal distribution. Then, the graphical device is a plot of $F^*_S(s_j) - F_n(s_j) $ versus $s_j$,
where $F^*_S(\cdot)$ is the maxentropic (cumulative) distribution function, $F_n(\cdot)$ is the empirical (cumulative) distribution function of
the observations and $\{s_j| j=1,...,n\}$ are sample data points of the total losses $S$. Under the hypothesis of marginal calibration, we expect minor
fluctuations about
zero. The same information can be visualized in terms of quantiles,

$$Q(F_S^*(\cdot) , q) - Q(F_n(\cdot), q), \quad q \in (0, 1)$$

\noindent
of the functions $F^*_S(\cdot)$ and $F_n(\cdot)$,  (Gneiting et. al, 2007).

%\bibliography{bsample}

\begin{thebibliography}{99}
\bibitem{} Barzilai, J., \& Borwein, J. M. (1988). {\it Two-point step size gradient methods}. IMA Journal of Numerical Analysis, {\bf 8}, 141-148.
\bibitem{} Berkowitz, J. (2001). {\it Testing density forecasts, with applications to risk management}. Journal of Business \& Economic Statistics, {\bf 19}, 465-474.
\bibitem{} Borwein, J.M. and Lewis, A.S (2000) {\em Convex Analysis and Nonlinear Optimization} CMS Books, Springer Verlag, New York, 2000.
\bibitem{} Clements, M. P., \& Hendry, D. F. (Eds.). {\it A companion to economic forecasting}. John Wiley \& Sons, (2008).
\bibitem{} Crow, E. L., \& Shimizu, K. (1988). {\it Lognormal distributions: Theory and applications} (Vol. 88). New York: M. Dekker.
\bibitem{} Cruz, M. G. (2002). {\it Modeling, measuring and hedging operational risk} (pp. 19-2). New York: John Wiley \& Sons.
\bibitem{} Diebold, F. X., Gunther, T. A. and Tay, A. S. (1998) {\it Evaluating density forecasts with applications
to financial risk management}, International Economic Review, {\bf 39}, 863-883.
\bibitem{} Gel, Y. R., \& Gastwirth, J. L. (2008). {\it A robust modification of the Jarque–Bera test of normality}. Economics Letters, 99(1), 30-32.
\bibitem{} Gneiting, T., Balabdaoui, F., \& Raftery, A. E. (2007). {\it Probabilistic forecasts, calibration and sharpness}. Journal of the Royal Statistical Society: Series B (Statistical Methodology), 69(2), 243-268.
\bibitem{} Gzyl, H., Novi-Inverardi, P.L. and Tagliani, A. (2013) {\it A comparison of numerical approaches to determine the severity of losses}. Journal of Operational Risk, {\bf 8} pp.3-15.
\bibitem{}Gzyl, H., \& Tagliani, A. (2012). Determination of the distribution of total loss from the fractional moments of its exponential. Applied Mathematics and Computation.
\bibitem{} Gzyl, H. and Vel\'asquez, Y. {\it Linear Inverse Problems: The maximum entropy connection}, World Scientific Pubs, Singapore, (2011).
\bibitem{} Hyndman, R. J., \& Koehler, A. B. (2006). {\it Another look at measures of forecast accuracy. International journal of forecasting}, {\it 22}(4), 679-688.
\bibitem{} Jaynes,E.T.  Information theory and statistical physics. {\em The Phys. Rev.}, {\bf 1957}{\em 106},  620-630.
\bibitem{} Leipnik, R.B. (1991) {\it On Lognormal Random Variables: I – The Characteristic Function}, Journal of the Australian Mathematical Society Series B, {\bf 32} pp.327–347.
\bibitem{} Kapur, J. N. (1989). {\it Maximum-entropy models in science and engineering}. John Wiley \& Sons, (1989)
\bibitem{} Marsaglia, G., \& Marsaglia, J. (2004). {\it Evaluating the anderson-darling distribution}. Journal of Statistical Software, 9(2), 1-5.
    \bibitem{} Panjer, H. {\em Operational Risk: Modeling and Analytics} John Wiley \& Sons,  New York, USA 2006.
\bibitem{} Rockafellar, R.T, and Uryasev, S. (2000) {\it Optimization of conditional value at risk}, Journal of Risk, {\bf 2}, 21-41.
\bibitem{} Rosenblatt, M. (1952). {\it Remarks on a multivariate transformation}, The Annals of Mathematical Statistics, {\bf 23} 470-472.
\bibitem{} Stephens, M. A. (1974). {\it EDF statistics for goodness of fit and some comparisons}, Journal of the American statistical Association, {\bf 69}, 730-737.
\bibitem{} Thas, O.  {\it Comparing distributions}. Springer, Berlin, (2010).
\bibitem{} Tay, A. S., and Wallis, K. F.(2000){\it Density forecasting: A survey}. Journal of Forecasting, {\bf 19},  235-254.


%Reprinted in Clements, M. P. and Hendry, D. F. (eds.) A Companion to Economic Forecasting,28
%pp.45 - 68, Oxford: Blackwells (2002).




\end{thebibliography}

\end{document}